\documentclass[12pt]{iopart}
\usepackage{graphicx}
\usepackage{iopams}
\usepackage{setstack}
\usepackage{amsfonts,amssymb,graphicx}
\usepackage{mathrsfs}
\usepackage{amsthm}
\usepackage[usenames]{color}
\usepackage{soul}
\usepackage{ulem}

\begin{document}
    \newcommand{\eg}{\textit{e.g.}{~}}
    \newcommand{\vphi}{\varphi}
    \newcommand{\bq}{\begin{equation}}
    \newcommand{\ba}{\begin{eqnarray}}
    \newcommand{\eq}{\end{equation}}
    \newcommand{\ee}{\end{equation}}
\newcommand{\ea}{\end{eqnarray}}
\newcommand{\tchi} {{\tilde \chi}}
\newcommand{\tA} {{\tilde A}}
\newcommand{\sech} { {\rm sech}}
\newcommand{\pstar}{\mbox{$\psi^{\ast}$}}
\newcommand {\tu} {{\tilde u}}
\newcommand {\tv} {{\tilde v}}
\newcommand{\dq}{{\dot q}}

\title[]{Soliton dynamics and stability in the ABS spinor model with a $\mathcal{PT}$-symmetric periodic complex  potential}

\author{Franz G.  Mertens$^1$ , Bernardo Sánchez-Rey$^2$ and Niurka R. Quintero$^3$}
\address{$^1$ Physikalisches Institut, Universit\"at Bayreuth, D-95440 Bayreuth, Germany}
\address{$^2$ Department of Applied Physics I, 
Escuela Politécnica Superior, University of Seville, 41011 Seville,
Spain}
\address{$^3$ Department of Applied Physics I, 
Escuela Técnica Superior de Ingeniería Informática, University of
Seville, Av. Reina Mercedes s/n,  41012 Seville, Spain}
\eads{\mailto{franzgmertens@gmail.com},\mailto{bernardo@us.es},\mailto{niurka@us.es}}
\date{\today}

\vspace{10pt}



\begin{abstract}

 We investigate the effects on solitons dynamics of introducing a
 $\mathcal{PT}$-symmetric complex potential in a specific family of the cubic Dirac equation in
(1+1)-dimensions,  called the ABS model. The potential is introduced
taking advantage of the fact that the nonlinear Dirac equation admits
a Lagrangian formalism. As a consequence, the imaginary part of the
potential, associated with gains and losses, behaves as a spatially periodic
damping (changing from positive to negative, and back) that acts at the same time on the two
spinor components.  A  collective coordinates theory is developed by making an ansatz for a moving soliton where the position, rapidity, momentum,
frequency, and phase are all functions of time. We consider the
complex potential as a perturbation and verify  that numerical solutions of
the equation of motions for the collective coordinates are in agreement
with simulations of the nonlinear Dirac equation. The main effect of
the imaginary part of the potencial is to induce oscillations in the
charge and energy (they are conserved for real potentials) with the
same frequency and phase as the momentum.  We find long-lived solitons even with
very large charge and energy oscillations. Additionally, we extend to the
nonlinear Dirac equation an
empirical stability criterion, previously employed successfully in the
nonlinear Schrödinger equation. 
\end{abstract}

\submitto{\JPA}

\vspace{2pc}

\maketitle

\section{\label{sec1}Introduction}

In quantum mechanics it is well known that  non-Hermitian but
$\mathcal{PT}$-symmetric Hamiltonians have  real eigenvalues. These Hamiltonians are
invariant  under the  simultaneous action of parity $\mathcal{P}$: $x
\to -x$ and time-reversal $\mathcal{T}$: $t \to -t$ operators, and can
be implemented by complex extensions of real Hamiltonians,   simply demanding that the complex potential
$U(x)$ satisfies $U(x)=U^*(-x)$ \cite{bender:1998,bender:2002,bittner:2004,bender:2007}.

Optical lattices provide an excellent test bench for these
$\mathcal{P} \mathcal{T}$ symmetric systems, due to the fact that the
Schrödinger equation with a complex potential is formally identical  to
the paraxial optical wave equation. The optical implementation   of
such systems requires a symmetric profile for the refractive index,
which plays the role of the real part of the potential, combined with
a properly designed distribution of gain and loss elements that models the
imaginary part of the potential \cite{guo:2009,ruter:2010}.   Soon, it was found that
$\mathcal{P} \mathcal{T}$-symmetric nonlinear optical 
lattices support soliton solutions, which turned out to be stable over a wide range
of parameters \cite{musslimani:2008,makris:2008,konotop:2016}. Interestingly, unlike generic dissipative systems, the
use of the $\mathcal{P} \mathcal{T}$ symmetry allows solitons to arise
in continuous families \cite{konotop:2014}, emulating in this sense 
localized modes transmission  in conservative
waveguides. These findings triggered numerous studies on the
interplay between nonlinearity and
$\mathcal{PT}$-symmetry \cite{kevrekidis:2015}. Particularly, new solitary waves were found
in a great variety of $\mathcal{P} \mathcal{T}$-symmetric waveguides arrays: dimers, that is,  a gain waveguide
linearly coupled with  a loss  waveguide, and Kerr nonlinearity in
each one \cite{sukhorukov:2010, dmitriev:2010}; chains of dimers or more complicated  $\mathcal{P}
\mathcal{T}$-units (oligomers) \cite{suchkov:2011,alexeeva:2012,li:2011,konotop:2012}; arrays with nonlinear gain-loss
terms \cite{abdullaev:2011};  lattices
with losses and gains localized in space ($\mathcal{P}
\mathcal{T}$-symmetric defects) \cite{hu:2011,shi:2011};  $\mathcal{P} \mathcal{T}$-symmetric optical
media with quadratic $\chi^{(2)}$ nonlinearity \cite{moreira:2012};    or even solitons in
the $\mathcal{PT}$-symmetric Gross-Pitaevskii  equation \cite{barashenkov:2016}.

Since a dimer, which can be considered the simplest $\mathcal{PT}$-symmetric
system, is described by two coupled nonlinear Schrödinger (NLS) equations,  a natural
step was to extend the gain and loss terms to the $(1+1)$-dimensional nonlinear
Dirac (NLD) equation \cite{cuevas:2016a,sakaguchi:2016}.  There is not an unique way to do that. In fact, in
Ref. \cite{alexeeva:2019} different $\mathcal{PT}$-symmetric
perturbations of the Dirac equation were discussed, and a
particular variant that preserves the invariance under Lorentz transformations and
$U(1)$ rotations was selected. Exact analytical soliton solutions were found in
the integrable $\mathcal{PT}$-symmetric Tirring model, which is a
completely integrable system with infinitely many conserved quantities, in
the $\mathcal{PT}$-symmetric Gross-Neveu model, where soliton energy
and momemtum are conserved,  and  in the $\mathcal{PT}$-symmetric
Alexeeva-Barashenkov-Saxena (ABS) model, where it was conjectured that
there are no conservation laws. Remarkably, in  these three
different families of the NLD equation, the spinor solitons 
turn out to be stable in most of their parameter domain.

In this paper, we essay another way to introduce
$\mathcal{PT}$-symmetric  terms into the NLD equation,
taking advantage of the fact that it admits a
Lagrangian. Specifically, we focus on the ABS
 model,  where recently we have
investigated soliton dynamics under ramp, harmonic, and periodic
real-potentials  $V(x)$ \cite{mertens:2021}. Soliton dynamics in the ABS
model turned out  to be qualitatively the
same as the dynamics of the Gross-Neveu soliton.  We now, in
accordance with the recipe to implement $\mathcal{PT}$-symmetric
Hamiltonians in quantum mechanics,   consider a complex extension $V(x)+i\,W(x)$,
such that $V(x)+i\,W(x)=V(-x)-i\,W(-x)$, being $V(x)$ and $W(x)$ real
functions. The imaginary part $W(x)$ of the potential is incorporated
into the ABS model through a dissipation function approach, similarly to
how a complex potential was introduced in a $\mathcal{PT}$-symmetric
NLS equation \cite{mertens:2016a}.  While the energy and the charge in the ABS model
with real potentials are conserved quantities, 
none of them are preserved  when introducing $W(x)$, because of  the
presence of dissipation.

To investigate soliton dynamics in the $\mathcal{PT}$-symmetric ABS model with complex
potential, we develop a
variational theory  using  an ansatz with five collective
coordinates, which was first introduced for the  Gross-Neveu
model under external or parametric driving
\cite{quintero:2019,quintero:2019a}. This collective coordinate (CC)
theory also allows us to study the robustness and 
stability of the solitary waves, by recovering an empirical stability
criterion successfully employed in the context of the NLS
equation. The criterion dictates that if the normalized momentum
 as a function of the soliton velocity, $\tilde{P}(v)$, 
 has a branch with negative slope, then the solitary wave is unstable. However, a positive slope of $\tilde{P}(v)$ does not guarantee stability.
The criterion was first enunciated to predict the instability of 
NLS solitons under spatio-temporal forces \cite{mertens:2011,cooper:2012}. 
It was subsequently also applied in the case of parametrically driven
solitons in the NLS equation \cite{mertens:2020}, and in the
generalized NLS equation \cite{quintero:2015}. In all these models,
when the CC formalism detected a branch with
negative slope in the $\tilde{P}(v)$ curve, the solitary wave turned
out to be unstable in simulations.

In Ref.\ \cite{mertens:2012},  it was
postulated a generalization of such criterion for the NLD
equation under potentials, as a sufficient dynamic condition
 for instability to arise.  
Nevertheless, it could not be verified neither for the external potentials 
studied in Refs.\
\cite{mertens:2012,nogami:1995,mertens:2016e,mertens:2021} nor for the
parametrically driven NLD equation \cite{cooper:2020},  because no
branch with negative slope for $\tilde{P}(v)$ was found in such
models, 
although there were instabilities reported  by numerical simulations. The generalization of the external potential to a complex
function, which preserves the  $\mathcal{PT}$-symmetry  in the ABS model, 
 has enabled 
 the validity of the empirical stability criterion in the NLD equations to be verified for the first time.

Our paper is organized as follows: in Section \ref{sec2}, we introduce
a complex potential in the ABS spinor model. The real part $V(x)$ is
added to the unperturbed Lagrangian density by using  the gauge
covariant derivative, while for the imaginary part  $W(x)$, a
dissipation functional approach is taken.  As a simple choice, 
$V(x)$ and $W(x)$ are considered to be periodic 
functions with, in principle, different periods and amplitudes.  Subsequently, in  Section \ref{sec3},  
we used a variational approach with an ansatz with five collective coordinates, that describe the position $q(t)$, rapidity $\beta(t)$,  momentum $p(t)$, frequency $\omega(t)$, and phase $\phi(t)$ of the soliton. 
We find that neither the energy, nor the momentum, nor the
charge are conserved quantities. This is the reason why it is
necessary to use so many collective variables. By using the Lagrangian
formalism with the dissipation function, we derive the equations of
motion satisfied by the collective variables, and obtain
analytical approximations for the soliton charge and energy, valid for small
amplitudes of $W(x)$  in the
non-relativistic regime. 
Simulations of the ABS model under complex potential are performed in Section \ref{sec4}, where the empirical stability criterion is verified for all considered sets of parameters. Finally, in Section \ref{sec5}, the main results  of the current work are summarized.

\section{ABS spinor model with a complex potential} \label{sec2}
The ABS model introduced in \cite{alexeeva:2019} belongs to  a family of the
cubic Dirac equation  in ($1+1$)-dimensions. It has the form
\begin{eqnarray}
	\label{eq2a}
	i\,(u_t-u_x)+v+u^\star\,v^2&=&0, \\
	\label{eq2b}
	i\,(v_t+v_x)+u+v^\star\,u^2&=&0, 
\end{eqnarray}
where $u(x,t)$ and $v(x,t)$ are the two components of the Dirac spinor
\begin{equation} \label{eq1}
	\Psi(x,t)=
	\left(\!
	\begin{array}{c}
		u(x,t) \\
		v(x,t)
	\end{array}
	\!\right).
\end{equation}
The system (\ref{eq2a})-(\ref{eq2b}) admits an exact explicit
solitary-wave solution, and in covariant  notation can be rewritten as
\begin{eqnarray}
	\label{eq4}
	i\,\gamma^{\mu}\,\partial_{\mu}\,\Psi+\Psi+(\bar{\Psi}\,\Psi)\,\Psi-\frac{1}{2}\,J_{\mu}\,\gamma^{\mu} \Psi&=&0.
\end{eqnarray}
Here, $\bar{\Psi}=\Psi^\dagger  \gamma^0$ is the conjugate spinor,  $\gamma^{\mu}$  are the Dirac gamma matrices for which the following representation is used
\begin{eqnarray} \label{q1}
	\gamma^{0} =
	\left( \begin{array}{cc}
		0 & 1  \\
		1 & 0  \end{array} \right), \qquad \gamma^{1} =
	\left( \begin{array}{cc}
		0 & 1  \\
		-1 & 0  \end{array} \right)\,,
\end{eqnarray}
and  $J_{\mu}=\bar{\Psi}\,\gamma_{\mu}\,\Psi$, with $\gamma_0=\gamma^{0}$ and $\gamma_1=-\gamma^{1}$.

 The nonlinear Eqs. (\ref{eq2a})-(\ref{eq2b}) for the
Dirac spinor can be derived from the Lagrangian density
\begin{eqnarray}
    \mathcal{L}_0 &=& \frac{i}{2}\,\Bigl[\bar{\Psi}\,\gamma^{\mu}\,\partial_{\mu}\,\Psi-\partial_{\mu}\,
    \bar{\Psi}\,\gamma^{\mu}\,\Psi\Bigr]+\bar{\Psi}\,\Psi+\frac{1}{2}\,(\bar{\Psi}\,\Psi)^2-\frac{1}{4}\,J_{\mu}\,J^{\mu} \nonumber \\
    &=&
    \frac{i}{2} \Bigl[ (u_t-u_x)\,u^\star -(u_t^\star-u_x^\star)\,u+
    (v_t+v_x)\,v^\star -(v_t^\star+v_x^\star)\,v \Bigr] + \nonumber \\
    &+& u\,v^\star + u^\star\,v+\frac{1}{2} \left[(u\,v^\star)^2+(u^\star\,v)^2\right].
    \label{eq5}
\end{eqnarray}
We now want to add a complex potential to the ABS model. The
real part $V(x)$ can be introduced through the gauge covariant 
derivative $i\,\partial_{\mu}\,\Psi \to
(i\,\partial_{\mu}-e\,A_{\mu})\,\Psi,$
thus adding  an additional term to the Lagragian density 
\cite{mertens:2012,mertens:2016e}
\begin{equation}
    \mathcal{L}_{3} = -e\,\bar{\Psi}\,\gamma^{\mu}\,A_{\mu}\,\Psi.
    \label{eq23}
\end{equation}
Using the axial gauge $e\,A_0=V(x)$, $A_{1}=0$, we then have
\begin{equation}
   \mathcal{L}_{3} = -V(x)\,\Psi^{\dagger}\,\Psi=-V(x)\,\Bigl[|u(x,t)|^2+|v(x,t)|^2  \Bigr].
    \label{eq24}
\end{equation}
Since $\mathcal{L}_{3}$  has to be a real function, the imaginary
part of the potential, $i\,W(x)$, cannot be introduced in the same
way. If $W(x)$ were constant, the term $i\,W_0\, \gamma^{0} \Psi$
that should appear in the partial differential equation for the
spinor would resemble a damping term \cite{quintero:2019a}. Consequently, we introduce
the imaginary part of the potential $i\, W(x)$  by
using  the dissipation function \cite{mertens:2016}
\begin{eqnarray} \label{disipa}
\cal{F}&=& -i\,W(x)\,(\bar{\Psi}\gamma^{0} \partial_t \Psi-\partial_t\bar{\Psi}\gamma^{0}  \Psi) \nonumber\\
&=&-i\,W(x)\,[u\,u_t^\star-u^\star\,u_t+v\,v_t^\star-v^\star\,v_t].
\end{eqnarray}
In standard fashion, from the  Euler-Lagrange equations
corresponding to the new Lagrangian density, $\mathcal{{L}} =
\mathcal{L}_0+\mathcal{L}_{3}$, and the dissipation function
$\cal{F}$
\begin{eqnarray} \label{eq4a}
\frac{d\,}{dt}\,\frac{\partial {\mathcal L} \quad}{\partial
u_t^\star}+ \frac{d\,}{dx}\,\frac{\partial {\mathcal L}
\quad}{\partial u_x^\star}-
\frac{\partial {\mathcal L}}{\partial u^\star}&=& \frac{\partial {\mathcal F} \quad}{\partial u_t^\star}, \\
\label{eq4b} \frac{d\,}{dt}\,\frac{\partial {\mathcal L}
\quad}{\partial v_t^\star}+ \frac{d\,}{dx}\,\frac{\partial {\mathcal
L} \quad}{\partial v_x^\star}- \frac{\partial {\mathcal L}}{\partial
v^\star}&=& \frac{\partial {\mathcal F} \quad}{\partial v_t^\star},
\end{eqnarray}
 we obtain the partial differential equations for the spinor components in the complex potential $V(x)+i   \, W(x)$:
\begin{eqnarray}
    \label{eq2apot}
    i\,(u_t-u_x)+v+u^\star\,v^2-[V(x)+i\,W(x)]\,u&=&0,
    \\
    i\,(v_t+v_x)+u+v^\star\,u^2-[V(x)+i\,W(x)]\,v&=&0.
    \label{eq2bpot}
\end{eqnarray}
This system of equations is $\mathcal{PT}$-symmetric
  as long as  $V(x)$ and  $W(x)$ are an even and an odd function, respectively. 
In the usual  $\mathcal{PT}$-symmetric NLD approach, one of the spinor
components gains and the second losses energy at an equal rate.  Notice
that here the approach is different. The imaginary part of the potential, associated with gains
and losses, acts on the two spinor components at the same time. 
In the following, we make the simple choice:
\begin{eqnarray}
    \label{potV}
    V(x)&=& -V_0\,\cos(k\,x), \\
    \label{potW}
    W(x)&=& - W_0\,\sin(l\,x),
\end{eqnarray}
where the amplitudes of the potentials, $V_0$ and $W_0$, and their wavenumbers, $k$
and $l$, are  in principle different quantities. Therefore, the term $W(x)$
represents a periodically varied damping, changing from positive to
negative, and back, thus providing a balanced gain and loss of energy
in the system. 

\section{ Variational  theory} \label{sec3}

The exact traveling wave solution of the ABS model  was
found in \cite{alexeeva:2019}. Explicitly, the two spinor components
are represented by
\begin{eqnarray} \label{eq20aa}
u(x,t) &=& \e^{-\beta/2} \,a[\gamma (x-v_s t)]\,\e^{i\,\theta[\gamma (x-v_s t)]}\,\e^{-i\,\omega\,\gamma(t-v_s x)},\\
\label{eq20bb} v(x,t) &=& -\e^{\beta/2} \,a[\gamma (x-v_s
t)]\,\e^{-i\,\theta[\gamma (x-v_s t)]}\,\e^{-i\,\omega\,\gamma(t-v_s
x)},
\end{eqnarray}
where  $\gamma=\cosh(\beta)=1/\sqrt{1-v_s^2}$ is the
Lorentz factor, and the velocity $v_s$ is constant. Furthermore,
\begin{eqnarray}\label{eq10a}
    \theta(x)&=&-\arctan[\lambda\,\tanh(\kappa\,x)], \\
    \label{eq9}
    a^2(x)&=& \,
    \frac{2 (1-\omega) \sech^2(\kappa x)[1+\lambda^2\,\tanh^2(\kappa\,x)]}{1-6 \lambda^2\,\tanh^2(\kappa\,x)+\lambda^4\,\tanh^4(\kappa x)},
\end{eqnarray}
where $\lambda=\sqrt{(1-\omega)/(1+\omega)}$, and
$\kappa=\sqrt{1-\omega^2}$. Note that $\lambda$ and $\kappa$ depend
on the frequency $\omega$ which is the only parameter of the solution. In
particular, the charge density 
$\rho(x,t)=|u(x,t)|^2+|v(x,t)|^2$
is bell-shaped for $\frac{3}{4}\,\le\,\omega\,<\, 1$,
and has two humps  for
$\frac{1}{\sqrt{2}}\,<\,\omega\,<\frac{3}{4}$.

An approximate solution for a moving soliton subjected to the
complex potential $V(x)+i\,W(x)$ can be constructed using a  trial
wave function with the same functional form as the exact solution
(\ref{eq20aa})-(\ref{eq20bb}).
 The trial function will depend on time through the so-called collective coordinates (CCs) describing the position $q(t)$, rapidity $\beta(t)$,  momentum $p(t)$, frequency $\omega(t)$, and phase $\phi(t)$ of the soliton, that is, we replace in (\ref{eq20aa})-(\ref{eq20bb})
\begin{eqnarray}
v_s\,t &\to& q(t), \quad \beta \to \beta(t),  \quad
\gamma\,\omega\,v_s \to p(t), \\
\omega\,\gamma\,t &\to& \phi(t)+p(t)\,q(t), \quad \omega \to
\omega(t), \label{eq25}
\end{eqnarray}
in order to construct the trial wave function
\begin{eqnarray}
u(z) &=& \e^{-\beta/2}\,a(z)\,\e^{i\,\theta(z)}\,\e^{-i\,\phi+i\,p\,z/\cosh(\beta)} ,\label{eq26a} \\
v(z) &=&
-\e^{\beta/2}\,a(z)\,\e^{-i\,\theta(z)}\,\e^{-i\,\phi+i\,p\,z/\cosh(\beta)},
\label{eq26b}
\end{eqnarray}
where $z=\cosh[\beta(t)] (x-q)$, and the functions $a(z)$ and
$\theta(z)$ depend on the variable $\omega(t)$. 

 We now insert our ansatz into the Lagrangian densities
(\ref{eq5}) and (\ref{eq24}). Integrating over $x$ leads to the
following results:
\begin{eqnarray} \nonumber
L_0&=&\int_{-\infty}^{+\infty} dx\, {\cal{L}}_0
=Q\,\Bigl[\dot{\phi}(t)+p\,\dot{q}-p\,\tanh(\beta)\Bigr]-
\frac{M_0(\omega)}{\cosh(\beta)}+\frac{I_2(\omega)}{\cosh(\beta)} \\ 
&-&I_0(\omega)\,\Bigl[\cosh(\beta)-\dot{q}\,\sinh(\beta)\Bigr],
\end{eqnarray}
and
\begin{eqnarray} \nonumber
    L_3&=&-\int_{-\infty}^{+\infty} dx\,
    V(x)\,\rho(x,t)=-2\int_{-\infty}^{+\infty}\,dz\,V\left(\frac{z}{\cosh \beta}+q\right)\,
    a^2(z).
\end{eqnarray}
where we have defined the integrals
\begin{eqnarray} \label{eq31}
I_0(\omega)&=&-2\int_{-\infty}^{+\infty} dz\,a^2(z)\,\theta'(z), \\
M_0(\omega)&=&2 \int_{-\infty}^{+\infty} dz\,a^2(z)\,\cos[2\,\theta(z)]=\sqrt{2}\,\ln\left[\frac{1+\sqrt{2}\,\kappa(\omega)}{1-\sqrt{2}\,\kappa(\omega)}\right],   \label{eq33} \\
\label{eq35} I_2(\omega)&=&\int_{-\infty}^{+\infty}
dz\,a^4(z)\,\cos[4\,\theta(z)].
\end{eqnarray}
In \cite{mertens:2021} it was found that these integrals
fulfill the relationships $I_0=I_2$, and $I_2=M_0-\omega\, Q$, where
the charge
\begin{eqnarray}
    \label{eq5da}
    Q(\omega)&=&\int_{-\infty}^{+\infty}\,dx\,\rho(x,t)
    = 2\,\ln\left[\frac{\omega+\kappa(\omega)}{\omega-\kappa(\omega)}\right],
\end{eqnarray}
and the rest mass $M_0$ are both functions of $\omega(t)$. Moreover, for the periodic potential given by 
Eq.\ (\ref{potV}), 
$L_3$ can be expressed as a periodic effective potential
\begin{eqnarray}
    L_3=-U(q,\beta,\omega)=V_0\,I_{4}(\beta,\omega)\,\cos(k\,q),
    \label{eq36}
\end{eqnarray}
with an amplitude modulated by the function
\begin{equation}
\label{eq35a}
I_4(\beta,\omega)=2\int_{-\infty}^{+\infty}\,dz\,a^2(z)\,\cos\left(\frac{k\,z}{\cosh
	\beta}\right).
\end{equation}
which depends on time through the rapidity and the frequency.
Therefore, the Lagrangian, in terms of the CCs, reads
\begin{eqnarray}
L&=&Q\,\Bigl[\dot{\phi}+p\,\dot{q}-p\,\tanh(\beta)\Bigr]-I_0\,\Bigl[\cosh(\beta)-\dot{q}\,\sinh(\beta)\Bigr]-\frac{\omega Q}{\cosh(\beta)}\nonumber \\
&& -U(q,\beta,\omega).\label{eq38}
\end{eqnarray}
The dissipation function in terms of the CCs is given by 
\begin{eqnarray}
\nonumber F &=&\int_{-\infty}^{+\infty} dx\,\mathcal{F}=
4\int_{-\infty}^{+\infty} dz\,
\widetilde{W}(z)\,a^2(z)\,\left[
\dot{\phi}-\frac{\dot{p}\,z}{\cosh(\beta)}+p\,\dot{q}\right]\\
\label{disipai} && \hspace*{-0.7cm}+
4\,\int_{-\infty}^{+\infty} dz\,
\widetilde{W}(z)
\,a^2(z)\,\left[\frac{\partial
\theta}{\partial z}\left(
\dot{\beta} \tanh^2(\beta)-\dot{q} \sinh(\beta)
\right)+\frac{\partial \theta}{\partial \omega}
\dot{\omega}\right],
\end{eqnarray}
where $\widetilde{W}(z)=W\left(z/\cosh(\beta)+q\right)$. 
For the particular choice of $W(x)$ specified in Eq.\ (\ref{potW}),
$F$ has the form
\begin{eqnarray}
\nonumber F&=& -2\,W_0\,\tanh(\beta)\,\cos(l\,q)\,
[\dot{\omega}\,I_6(\beta,\omega)-\tanh(\beta)\,\dot{\beta}\,I_5(\beta,\omega)] \\
&&-
 2\,W_0\,\sinh(\beta)\,\sin(l\,q)\,\dot{q}\,I_{7}(\beta,\omega)+
 2\,W_0\,\frac{\dot{p}\,\cos(l\,q)}{\cosh(\beta)}\,I_{8}(\beta,\omega) \nonumber\\
 &&
 -2\,W_0\,\sin(l\,q)\,(\dot{\phi}+p\,\dot{q})\,I_{9}(\beta,\omega), 
\end{eqnarray} 
where
\begin{eqnarray}
\label{i5}
I_5(\beta,\omega)&=&-2\,\int_{-\infty}^{+\infty}\,dz\,\sin\left(\frac{l\,z}{\cosh
\beta}\right)\,
z\,a^2(z)\,\frac{\partial\theta}{\partial z},\\
\label{i6}
I_6(\beta,\omega)&=&2\,\int_{-\infty}^{+\infty}\,dz\,\sin\left(\frac{l\,z}{\cosh
\beta}\right)\,
a^2(z)\,\frac{\partial\theta}{\partial \omega},\\
\label{i7}
I_7(\beta,\omega)&=&-2\,\int_{-\infty}^{+\infty}\,dz\,\cos\left(\frac{l\,z}{\cosh
\beta}\right)\,
a^2(z)\,\frac{\partial\theta}{\partial z},\\
\label{i8}
I_8(\beta,\omega)&=&2\,\int_{-\infty}^{+\infty}\,dz\,\sin\left(\frac{l\,z}{\cosh
\beta}\right)\,
z\,a^2(z),\\
\label{i9}
I_9(\beta,\omega)&=&2\,\int_{-\infty}^{+\infty}\,dz\,\cos\left(\frac{l\,z}{\cosh
\beta}\right)\,a^2(z).
\end{eqnarray}
Once $L$ and $F$ have been obtained as functions of
the CCs, we are ready to derive equations of motion for the CCs.

\subsection{Equations of motion}

 From the Lagrange equation for $\phi(t)$,
\begin{equation}
\frac{d}{dt}\frac{\partial L}{\partial \dot{\phi}}=\frac{\partial
L}{\partial \phi}+\frac{\partial F}{\partial \dot{\phi}}, \label{lphi}
\end{equation}
the first equation of motion reads
\begin{equation}
\label{eqomega}
\frac{dQ}{d\omega} \, \dot{\omega}=-2\,W_0\,\,I_9(\beta,\omega)\, \sin(l\,q).
\end{equation}
From the Lagrange equation for $p(t)$,
\begin{equation}
\frac{d}{dt}\frac{\partial L}{\partial \dot{p}}=\frac{\partial
L}{\partial p}+\frac{\partial F}{\partial \dot{p}}, \label{lp}
\end{equation}
a new relationship between the Lorentz factor and the rapidity is
given by
\begin{equation}
\label{eqq}
\dot{q}=\tanh(\beta)-\frac{2\,W_0\,\cos(l\,q)}{\cosh(\beta) Q(\omega)}\,I_8(\beta,\omega).
\end{equation}
Only if  $W_0=0$, $\dot{q}=\tanh(\beta)$, and therefore $\gamma=\cosh(\beta)$.

From the Lagrange equation for $\beta(t)$,
\begin{equation}
    \frac{d}{dt}\frac{\partial L}{\partial \dot{\beta}}=\frac{\partial L}{\partial \beta}+\frac{\partial F}{\partial \dot{\beta}},
    \label{lbeta}
\end{equation}
an algebraic equation is obtained that links $p(t)$ with $\omega(t)$, $\beta(t)$, and $q(t)$.
\begin{eqnarray}
    \nonumber 
        &&p=\omega\,\sinh(\beta)-\frac{2\,W_0\,\cos(l\,q)}{Q^2}\,I_8(\beta,\omega)\,I_0(\omega)\,\cosh^2(\beta) \\ \label{eqbeta}&-&\frac{\partial U}{\partial \beta}\,\frac{\cosh^2(\beta)}{Q}+\frac{2\,W_0}{Q}\,\cos(l\,q)\,\sinh^2(\beta)
    \,I_5(\beta,\omega).
\end{eqnarray}
From the Lagrange equation for $q(t)$,
\begin{equation}
\frac{d}{dt}\frac{\partial L}{\partial \dot{q}}=\frac{\partial
L}{\partial q}+\frac{\partial F}{\partial \dot{q}}, \label{lq}
\end{equation}
an evolution equation  is obtained  for the canonical momentum 
\begin{equation}
	P_q=\frac{\partial L}{\partial \dot{q}}=p\,Q(\omega)+\sinh(\beta)\,I_0(\omega), \label{eqPq}
\end{equation}
which agrees with the field momentum 
\begin{equation}
	P=\int_{-\infty}^{+\infty}\,dx\,\mathcal{P}(x,t)=\int_{-\infty}^{+\infty} dx\, \frac{i}{2}\,[u_x^\star\,u-u_x\,u^\star+v_x^\star\,v-v_x\,v^\star].  \label{eqP}
\end{equation}
The momentum $P$ is governed by
\begin{equation}
\label{eqp} \dot{P}=-\frac{\partial U}{\partial
q}-2\,W_0\,\sin(l\,q)[p\,I_9(\beta,\omega)+\sinh(\beta)\,I_7(\beta,\omega)].
\end{equation}
Finally,  from the Lagrange equation for $\omega(t)$
\begin{equation}
\frac{d}{dt}\frac{\partial L}{\partial \dot{\omega}}=\frac{\partial
L}{\partial \omega}+\frac{\partial F}{\partial \dot{\omega}},
\label{lomega}
\end{equation}
an independent equation of motion for the phase $\phi(t)$ is obtained:
\begin{eqnarray}
	&&\frac{dQ}{d\omega}\left[\dot{\phi}+p\Bigl(\dot{q}-\tanh(\beta)\Bigr)-\frac{\omega}{\cosh
		(\beta)}\right]-Q\,\sinh(\beta)\,\Bigl[\dot{q}-\tanh(\beta)\Bigr]
	\nonumber \\
	&& \label{eqomega1} -\frac{\partial
		U}{\partial \omega}= 2\,W_0\,\cos(l\,q)\,\tanh(\beta) \,I_6(\beta,\omega).
\end{eqnarray}
By using the  equations of motion of the CCs (\ref{eqomega}), (\ref{eqq}), (\ref{eqbeta}), (\ref{eqp}), and (\ref{eqomega1}) we can also derive an expression for the soliton energy in terms of the CCs. The energy density is
\begin{eqnarray}
	T^{00}&=&
	\frac{i}{2}\left[u_xu^{\star}-u_x^{\star}u-v_xv^{\star}+v_x^{\star}v \right]
	-(uv^{\star}+u^{\star}v) \nonumber
	\\ 
	&&-\frac{1}{2}\left[(uv^{\star})^2+(u^{\star}v)^2 \right]
	+V(x)\left[|u|^2+|v|^2\right]. \label{eqT00}
\end{eqnarray}
Therefore,
\begin{equation}
	\label{eqE}
	E=\int_{-\infty}^{+\infty}dx\,T^{00}(x,t). 	
\end{equation} 
Inserting the ansatz (\ref{eq26a})-(\ref{eq26b}) in the above equation
and by integrating over $x$, we obtain
\begin{equation}
	E= I_0(\omega) \cosh(\beta)+p  Q(\omega)
        \tanh(\beta)+\frac{\omega\,
          Q(\omega)}{\cosh(\beta)}+U(q,\beta,\omega).  \label{eqECC}	
\end{equation} 
The energy in this system  is not a conserved quantity because 
the dissipation function acts as a source term in the continuity equation for the energy density
\begin{equation}
	\label{eqT00c}
	T_{t}^{00}+{\cal{J}}_x={\cal{F}},
\end{equation}
where the energy current density ${\cal{J}}$ is 
\begin{equation}
	\label{eqdcE}
	{\cal{J}}(x,t)=\frac{i}{2}[u_t^\star\,u-u_t\,u^\star+v_t\,v^\star-v_t^\star\,v].
\end{equation}
Hence, assuming ${\cal{J}}(\infty,t)-{\cal{J}}(-\infty,t)=0$, we obtain 
\begin{equation}
	\label{eqconE}
	\frac{dE}{dt}=F.
\end{equation}

\subsection{Nonrelativistic regime}

In order to study the influence of the imaginary part of the potential in the soliton dynamics, let us consider the limit in which $W_0$ is small and expand the CCs in powers of $W_0$ . For instance, for the soliton position we write: 
\begin{equation}
	q(t)=q^{(0)} (t)+ W_0 \, q^{(1)}(t)+ W_0^2 \, q^{(2)}(t)+\dots
	\label{eq:expand}
\end{equation}
At zero order, we have the system of equations: 
\begin{eqnarray}
	\dot{\omega}^{(0)}(t)&=&0, \label{eqwaprox}\\
	\label{eqqaprox}
	\dot{q}^{(0)}(t)&=&\tanh[\beta^{(0)}(t)], \\
	\label{eqpaprox}
	\dot{P}^{(0)}(t)&=&-\frac{\partial U}{\partial q^{(0)}}, \\
	\label{eqbetaaprox}
	 p^{(0)}(t)&=&\omega^{(0)}\,\sinh[\beta^{(0)}(t)]. 
\end{eqnarray}
Therefore, $\omega^{(0)}= \omega(0)$ is constant, and  $Q^{(0)}=Q[\omega(0)]$ is also constant since the charge only depends on the frequency. 

In the non-relativistic regime $|\dot{q}(t)| \ll 1$,
$\beta^{(0)}=\dot{q}^{(0)}$ and   $\cosh{\beta}^{(0)}\approx 1$.  In
this regime we can neglect the dependence on $\beta^{(0)}$ of the
integrals  $I_{i}[\beta^{(0)},\omega^{(0)}] \simeq
I_{i}[\omega^{(0)}]$. Consequently, from Eq. (\ref{eq36})  we obtain 
\begin{equation}
	\label{eq:particlepot}
	U[q^{(0)}(t)]=-V_0\,I_4[\omega^{(0)}]\,\cos[k\,q^{(0)}(t)], 
      \end{equation}
and  the system (\ref{eqwaprox})-(\ref{eq:particlepot}) 
leads to the pendulum equation \cite{mertens:2021}
\begin{eqnarray}
	\label{eq:pen}
	\ddot{q}^{(0)}(t)+\frac{k\,V_0\,I_4[\omega^{(0)}]}{M_0[\omega^{(0)}]}\sin[k\,q^{(0)}(t)]=0.
\end{eqnarray}
For an initial position $q^{(0)}(0)=2\,n\,\pi/k$ ($n \in Z$),  such
that the soliton center of mass is at the minimum of the potential,
and for a positive initial velocity $\dot{q}^{(0)}(0) > 0$, the solution
of the above equation is 
\begin{eqnarray}
	\label{eq:pos}
	q^{(0)}(t)&=&q^{(0)}(0)+\frac{2}{k} am\left(\frac{k\,\dot{q}^{(0)}(0)}{2}t,m_1\right)
\end{eqnarray}
if $m_1<1$, and 
\begin{eqnarray}
	\label{eq:pos1}	
	q^{(0)}(t)&=&q^{(0)}(0)+\frac{2}{k} \arcsin \left\{\frac{1}{m_1} \sin \left[am\left(\frac{k\,m_1\,\dot{q}^{(0)}(0)}{2}t,\frac{1}{m_1}\right)\right] \right\}
\end{eqnarray}
if $m_1>1$. The parameter $m_1=v_c/\dot{q}^{(0)}$ is the modulus in the Jacobi amplitude $am$,  and 
\begin{equation}
v_c=2\,\sqrt{\frac{V_0 I_4[\omega^{(0)}]}{M_0[\omega^{(0)}]}},
\end{equation}
is a critical velocity 
such that for $\dot{q}^{(0)}(0)<v_c$ ($m_1>1$), the soliton oscillates
around the minimum of the potential, whereas for $\dot{q}^{(0)}(0)>v_c$
($m_1<1$) the soliton motion is unbounded.
Moreover, note that from Eqs. (\ref{eqPq})  and (\ref{eqpaprox}), it can be easily deduced that
\begin{eqnarray}
	\label{eq:Psuper0}
	P^{(0)}(t)&=&M_0[\omega^{(0)} ] \,  \dot{q}^{(0)}(t)
                      =M_0[\omega^{(0)}] \, \dot{q}^{(0)}(0)\,dn\left[\frac{k\,\dot{q}^{(0)}(0)}{2}t,m_1\right], 
\end{eqnarray}
where $dn(\cdot,\cdot)$ is  the delta amplitude, which is one of the Jacobi elliptic functions.  Consequently, at zero order, the momentum is a periodic function with frequency
\begin{equation}
\Omega_{osc}=\frac{\pi}{4}  \frac{ k\,\dot{q}(0)\,m_1}{ K\left(1/m_1\right)} \label{eqT1}
\end{equation}
if $m_1>1$ (oscillatory motion), or  
\begin{equation} \label{eqT2}
  \Omega_{unb}=\frac{\pi}{2} \frac{ k\,\dot{q}(0)}{ K(m_1)} 
\end{equation} 
if $m_1<1$ (unbounded motion), where $K(m)=F(\pi/2,m)$ and
$F(y,m)=\int_{0}^{y} dz/\sqrt{1-m^2 \sin^2 z}$ is the elliptic
integral of the first kind.

At first order in $W_0$, the equations of motion for the CCs are extremely 
complicated  and it is therefore not worth including them here. Furthermore,  what we are
mainly interested in involves the first correction to the
charge and the energy, since both are constant at zero order.  From Eq.\ (\ref{eqomega}), the first-order correction to the charge 
satisfies
\begin{equation}
	\label{eqQ1}
	\frac{dQ^{(1)}}{dt}=-2\,I_9[\omega^{(0)}]\, \sin[l\,q^{(0)}(t)].
\end{equation}
In the case of $l=k$, $I_9[\omega^{(0)}]=I_4[\omega^{(0)}]$, and taking
into account Eqs. (\ref{eq:pen})  and  (\ref{eq:Psuper0}),  we obtain 
\begin{equation}
	\label{eq:Qleqk}
	\frac{dQ^{(1)}}{dt}
	=\frac{2\,M_0[\omega^{(0)}]}{k\,V_0}\,\ddot{q}^{(0)}(t)=\frac{2}{k\,V_0}\,
	\dot{P}^{(0)}, 
\end{equation}
which leads to
\begin{equation}
	\label{eq:Q1}
	Q^{(1)}(t)= \,\frac{2}{k\,V_0}\, [P^{(0)}(t)-P(0)]. 
\end{equation}
That is, whenever $l=k$, the complex part of the potential induces
oscillations in the charge  with the same frequency as the momentum.

From  Eq.\ (\ref{eqECC}),   the first-order correction to the  energy
is approximately  given by
\begin{equation}
	\label{eq:E1}
	E^{(1)}(t) \approx \omega^{(0)} \,  Q^{(1)}(t), 
\end{equation}
and hence we expect, at least in the case $l=k$, that the energy will oscillate
with the same frequency and in phase  with the charge and the
momentum.  

When  $ l \ne k$, we do not have an analytical approximation for
$Q^{(1)}(t)$ and $E^{(1)}(t)$, but looking at (\ref{eqQ1})  where one
finds  the sine of an oscillating function, it is expected that the
charge and the momentum will oscillate with multiple frequencies.

\section{Simulations} \label{sec4}

We now compare simulations of the ABS nonlinear spinor model with complex potential  (\ref{eq2apot})-(\ref{eq2bpot}) with numerical solutions of the equations of motion of the CCs (\ref{eqomega}), (\ref{eqq}), (\ref{eqbeta}), (\ref{eqp}), (\ref{eqomega1}), and also with the analytical approximations derived in the non-relativistic regime. For the numerical solution of the CC equations of motion we have to supply the initial conditions  $\omega(0)$, $q(0)$, and $\dot{q}(0)$,  and use  Eqs. (\ref{eqq}) and
(\ref{eqbeta}) to obtain $\beta(0)$ and $p(0)$, respectively. In all cases, we have additionally  taken $\phi(0)=0$. 

For the simulation of the spinor equations, we discretize the system by taking constant spatial intervals $\Delta x=0.02$, and a sufficiently large number of points $N$  such that the length of the system is 
considerably longer than the soliton width.  A Runge-Kutta-Verner fifth-order algorithm with variable time step, and a spectral method for the computation of spatial derivatives \cite{cuevas:2015} is employed to integrate over time. As initial condition, we take the initial values of the CCs and insert them in the trial wave function (\ref{eq26a})-(\ref{eq26b}) at $t=0$: 
\begin{eqnarray}
    \label{eq70a}
    u(x,0)&=& e^{-\beta(0)/2}\,\tilde{a}(x)\,e^{i\,\tilde{\theta}(x)}\,e^{i\,p(0)\,(x-q(0))},
    \\
    v(x,0)&=& -e^{\beta(0)/2}\,\tilde{a}(x)\,e^{-i\tilde{\theta}(x)}\,e^{i\,p(0)\,(x-q(0))},
    \label{eq70b}
\end{eqnarray}
where $\tilde{a}(x)=a(\cosh[\beta(0)][x-q(0)])$ and $\tilde{\theta}(x)=\theta(\cosh[\beta(0)][x-q(0)])$. 
Once we have computed  
the spinor components, the soliton position can be calculated by means of
\begin{equation}
    \label{eq71}
    q(t)=\frac{{\int\,dx\,x\,\rho(x,t)}}{\int\,dx\,\rho(x,t)}.
\end{equation}
Additionally, we compute the  soliton momentum given by Eq.~(\ref{eqP}),
and the soliton energy using (\ref{eqT00}) and (\ref{eqE}).   

\subsection{Case $l=k$}
To start with, we investigate the influence of $W_0$ on the critical
velocity $v_c$ that marks the transition from oscillatory to unbounded
soliton motion.  To this end, we fix the  values of $\omega(0)=0.9,
q(0)=0$, $l=k=\pi/32$, $V_0=0.001$, vary $W_0 \in [0,0.01]$, and look
for the initial velocity $\dot{q}(0)$ for which the soliton starts an
unbounded motion. We have chosen $k=\pi/32$ so that the
 spatial period of the potential $\lambda=2\pi/k=64$ is much larger
 than the width of the soliton. Moreover, this exact value of the spatial
 period allows us to use rigorously periodic boundary conditions in
 the simulations, by taking an integer multiple of $\lambda$ as the system length $L$. For other values of the wavelength, the results do not change qualitatively whenever the two length scales of the problem (spatial period and soliton width) remain sufficiently separated.

\begin{figure}[h!]
	\centering
	\begin{tabular}{c}
		\includegraphics[width=0.5\linewidth]{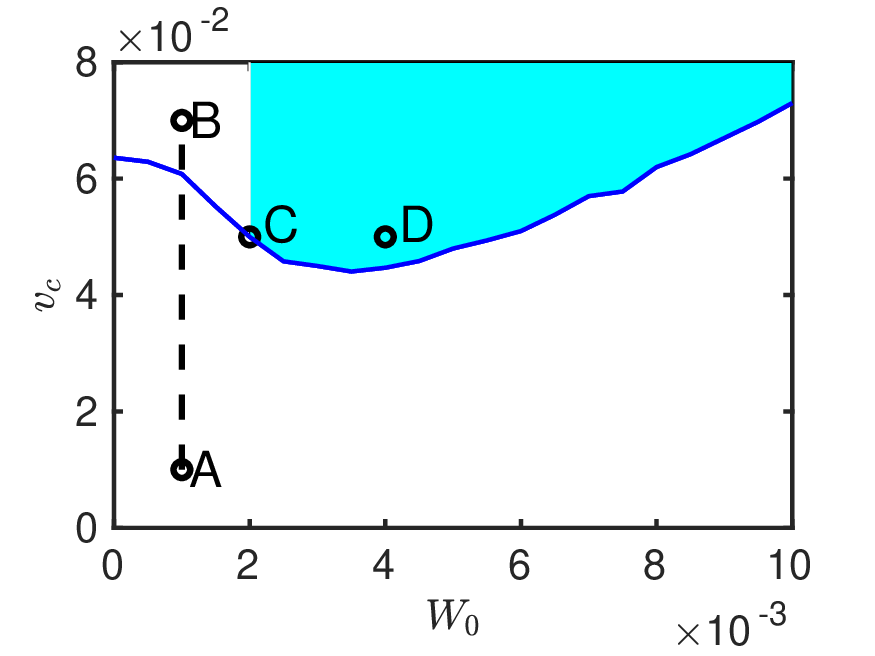} 
	\end{tabular}
	\caption{Critical velocity, $v_c$, versus amplitude, $W_0$, of
          the imaginary part of the potential, computed from the
          collective coordinate equations. For initial velocities
          above (below) $v_c$, the
          soliton motion is unidirectional (oscillatory). In the blue
          shaded region, the stability curve $\tilde{P}(\dot{q})$
          has at least one branch with  negative slope and thus instability is expected. The dashed
          line between A and B is
          a guide for the eyes. 	Parameters: $V_0=0.001$, $k=l=\pi/32$. Initial conditions: $\omega(0)=0.9$, $q(0)=0$, and $\phi(0)=0$. 
	}
	\label{fig1}
\end{figure}

\begin{figure}
\begin{center}
	\begin{tabular}{cc}
		\includegraphics[width=0.4\linewidth]{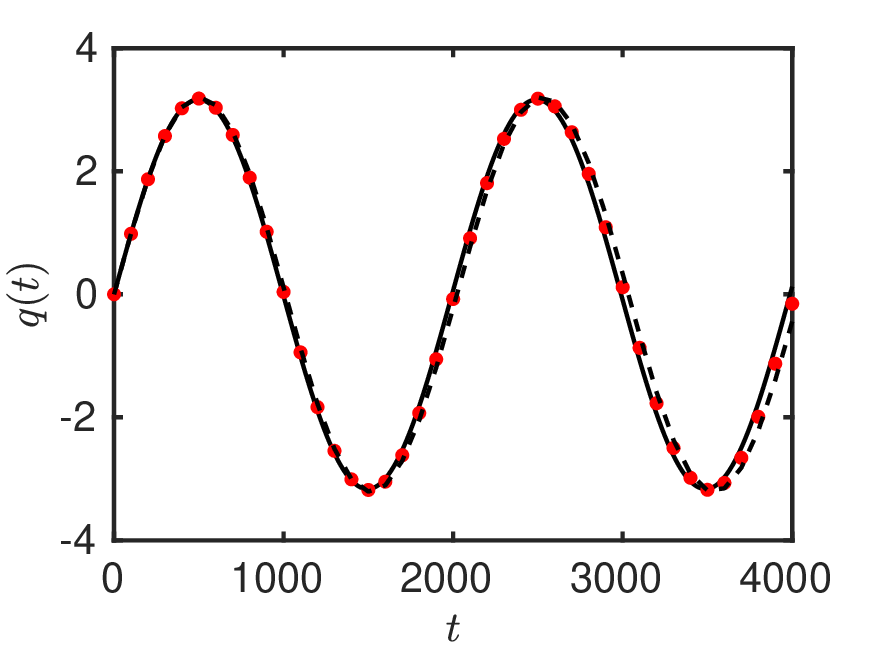} &  \includegraphics[width=0.4\linewidth]{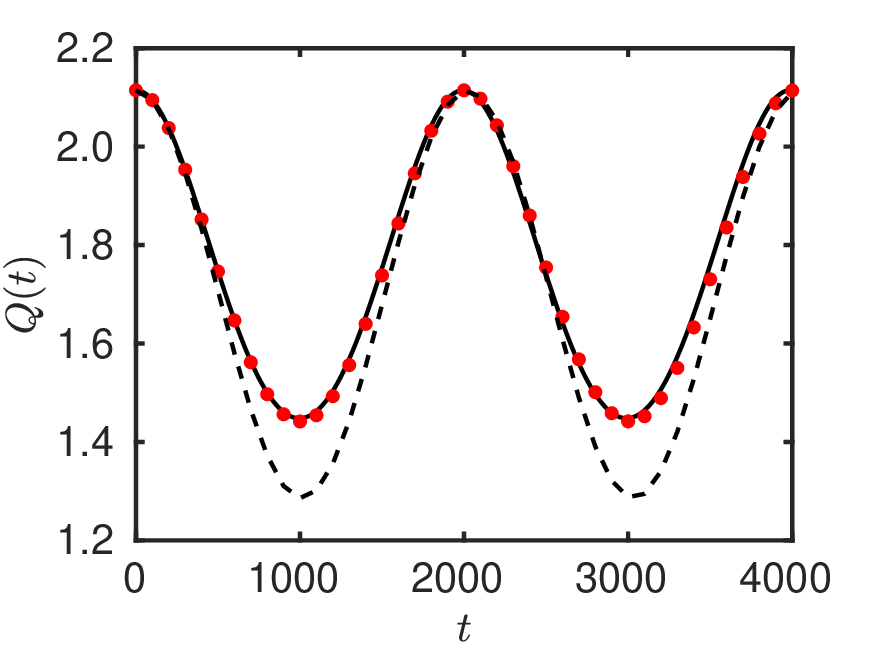} \\
		\includegraphics[width=0.4\linewidth]{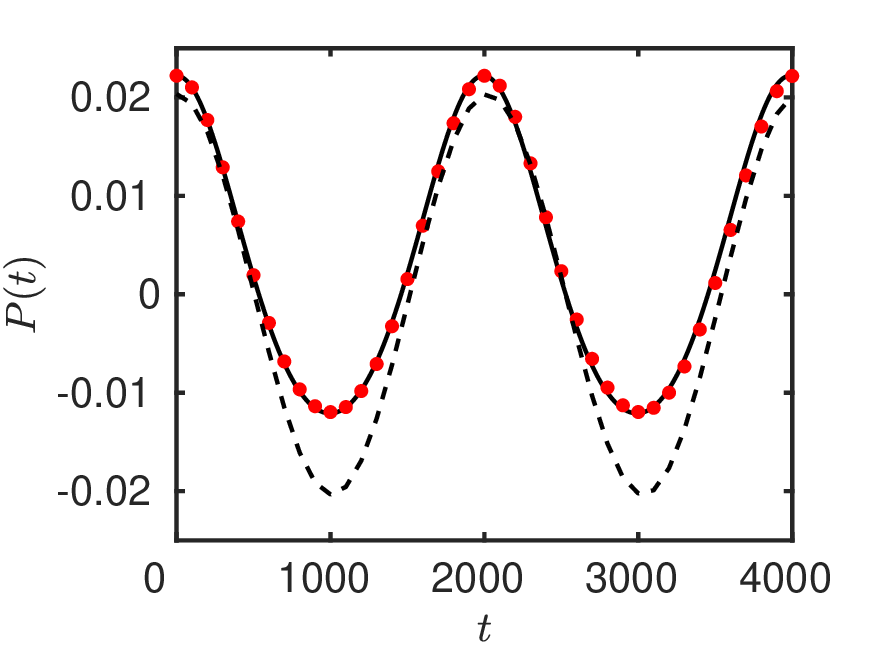} &  \includegraphics[width=0.4\linewidth]{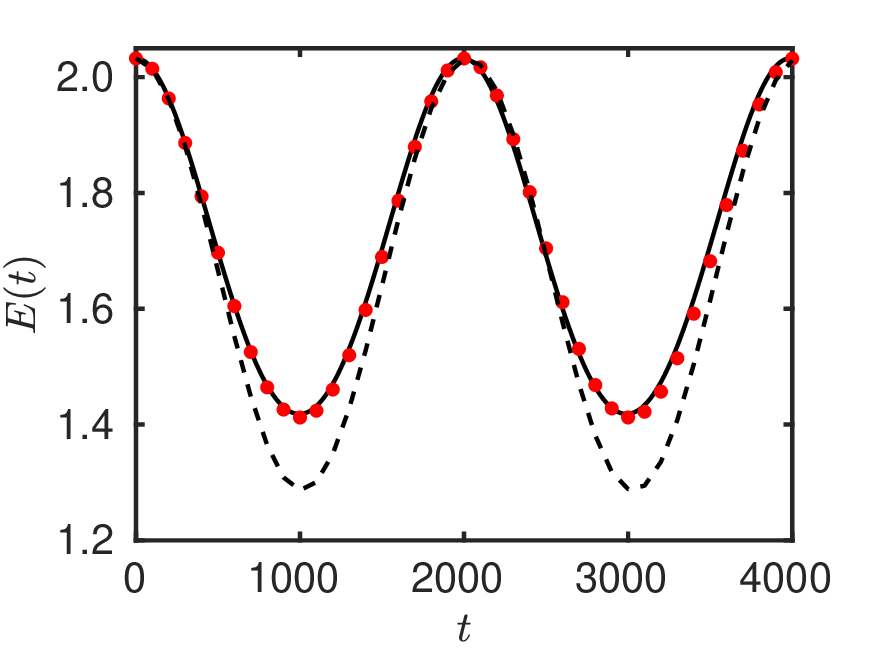}
	\end{tabular}
	\caption{
		Soliton oscillatory motion in the complex potential (\ref{potV})-(\ref{potW}) for  $V_0=W_0=0.001$, and $k=l=\pi/32$. Evolutions of $q(t)$, $P(t)$, $Q(t)$, and $E(t)$  are shown. 
		Solid lines and red points correspond to simulations  and collective coordinate results, respectively. Left-hand panels: dashed lines represent  approximate analytical solutions at zero-order  for $q(t)$ and $P(t)$, given by Eqs.~(\ref{eq:pos}) and (\ref{eq:Psuper0}), respectively. Right-hand panels: dashed lines show approximate analytical solutions for the charge  and the energy up to the first-order correction. Initial conditions: $\omega(0)=0.9$,
		$q(0)=0$, $\phi(0)=0$, and $\dot{q}(0)=0.01$.}
	\label{fig2}
\end{center}
\end{figure}

\begin{figure}
	\centering
	\begin{tabular}{cc}
		\includegraphics[width=0.4\linewidth]{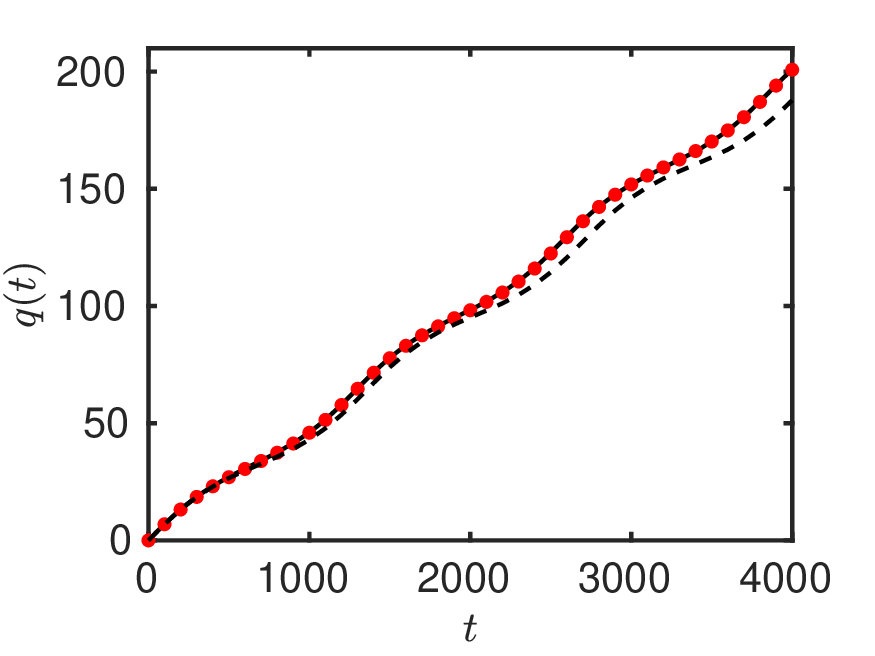} &     \includegraphics[width=0.4\linewidth]{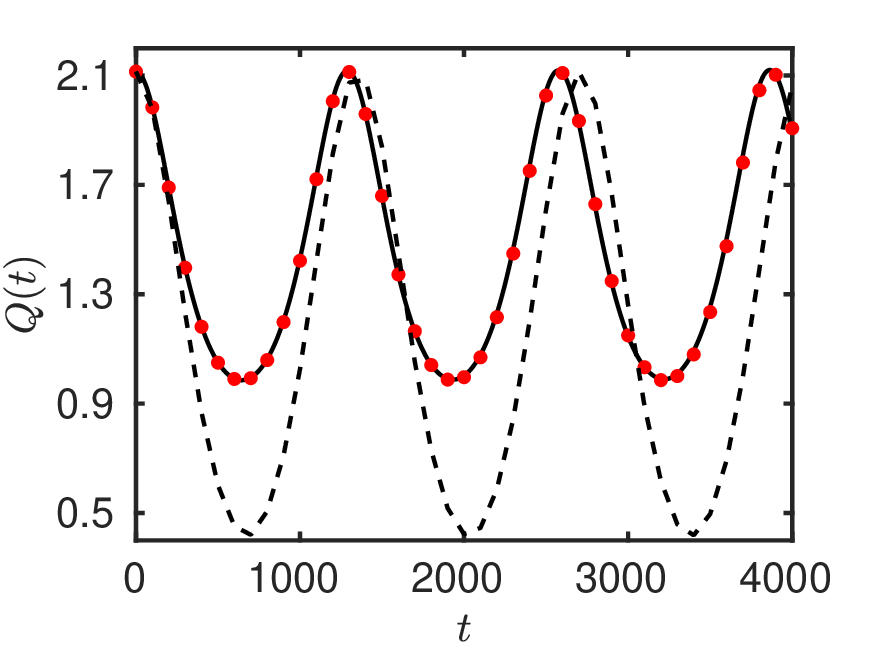} \\
		\includegraphics[width=0.4\linewidth]{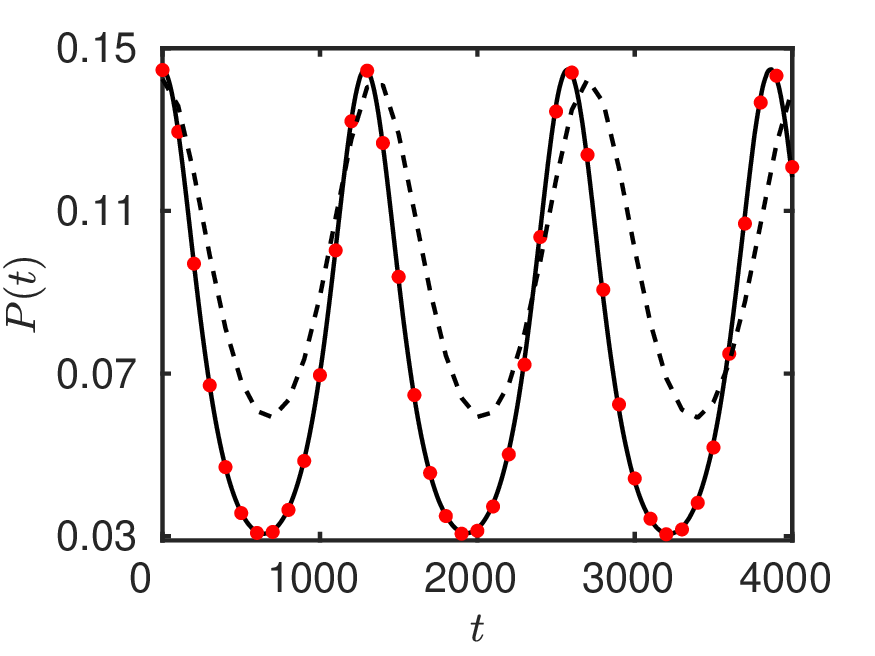} &  \includegraphics[width=0.4\linewidth]{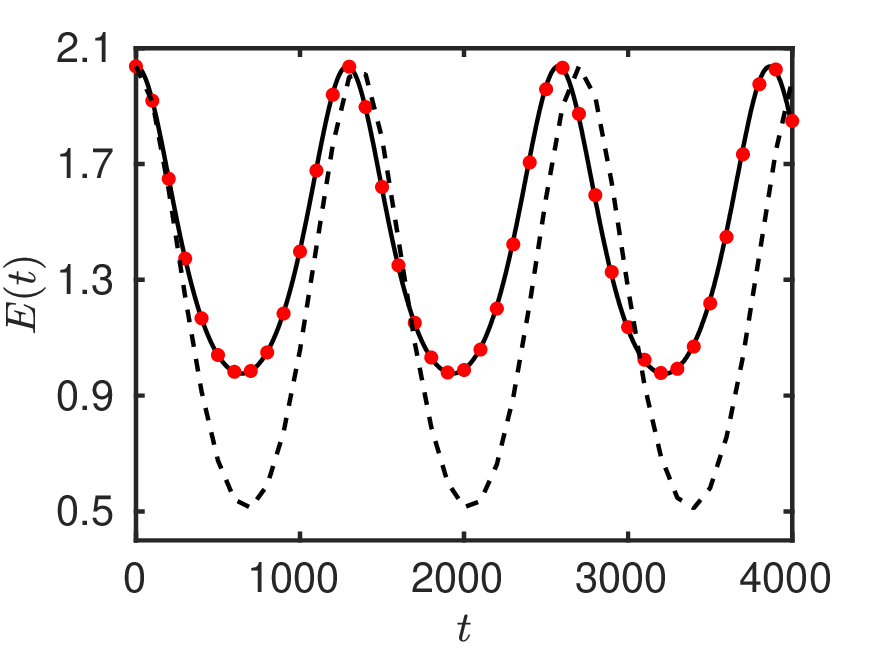}
	\end{tabular}
	\caption{
		Unbounded motion of the soliton in the complex potential (\ref{potV})-(\ref{potW}) for $V_0=W_0=0.001$, and $k=l=\pi/32$. Evolutions of $q(t)$, $P(t)$, $Q(t)$, and $E(t)$ are shown.
		Solid line and red points correspond to simulations and collective coordinate  results, respectively.  Left-hand panels: dashed lines represent approximate analytical solutions at zero-order for $q(t)$ and $P(t)$, given by 
		Eqs.~(\ref{eq:pos}) and (\ref{eq:Psuper0}), respectively. Right-hand panels: dashed lines show  
		approximate analytical solutions for the charge and the energy up to the first-order correction.
		Initial conditions: $\omega(0)=0.9$,
		$q(0)=0$, $\phi(0)=0$, and $\dot{q}(0)=0.07$.
	}
	\label{fig3}
\end{figure}

In  Fig.~\ref{fig1}, it is shown that  $v_c$ displays a minimum at $W_0=0.0035$ as $W_0$ increases.  Soliton dynamics below and above the critical velocity for $W_0=0.001$ (points A and B) are shown in Figs.~\ref{fig2} and
\ref{fig3}. Note the excellent agreement between simulations (solid line) and the CC theory (red points).
The dashed lines in the left-hand panels of these figures represent
the approximate analytical solutions at order zero $q^{(0)}(t)$ and
$P^{(0)}(t)$. 
Dashed lines are also used in the right-hand panels to 
represent the analytical approximations to the charge and the energy
up to the first-order corrections. Notice that, in accordance with the
approximations  (\ref{eq:Q1}) and (\ref{eq:E1}), the momentum, the
charge, and the energy all oscillate in phase with the same frequency,
given approximately by (\ref{eqT1})  in the case of oscillatory motion, or by
(\ref{eqT2}) in the case of unbounded motion. Logically, the analytical approximations work better for the oscillatory motion ($\dot{q}(0)<v_c$),  since they were derived in the non-relativistic regime $\dot{q}(t)\ll 1$. 

\begin{figure}
	\centering
	\begin{tabular}{cc}
		\includegraphics[width=0.4\linewidth]{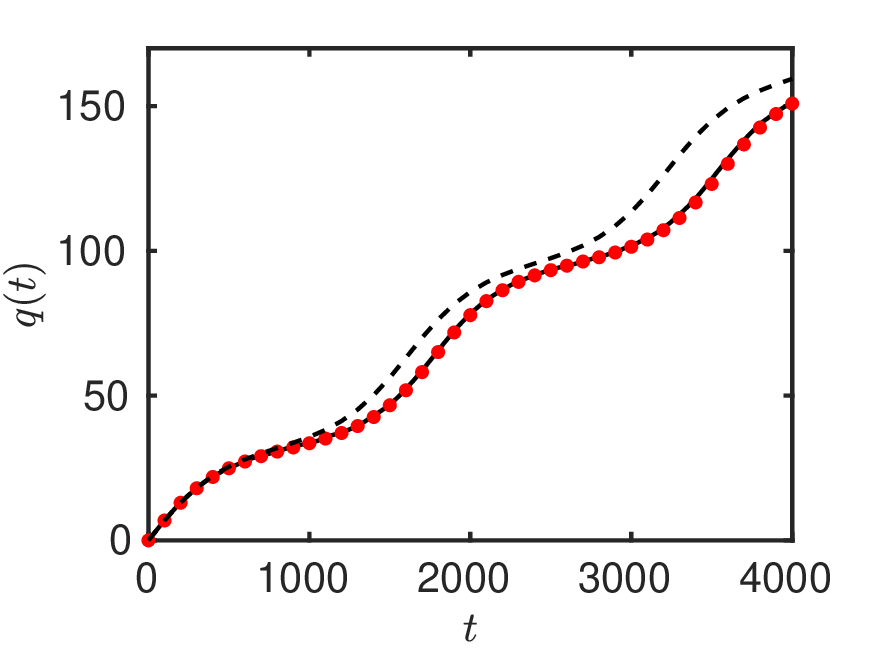} &     \includegraphics[width=0.4\linewidth]{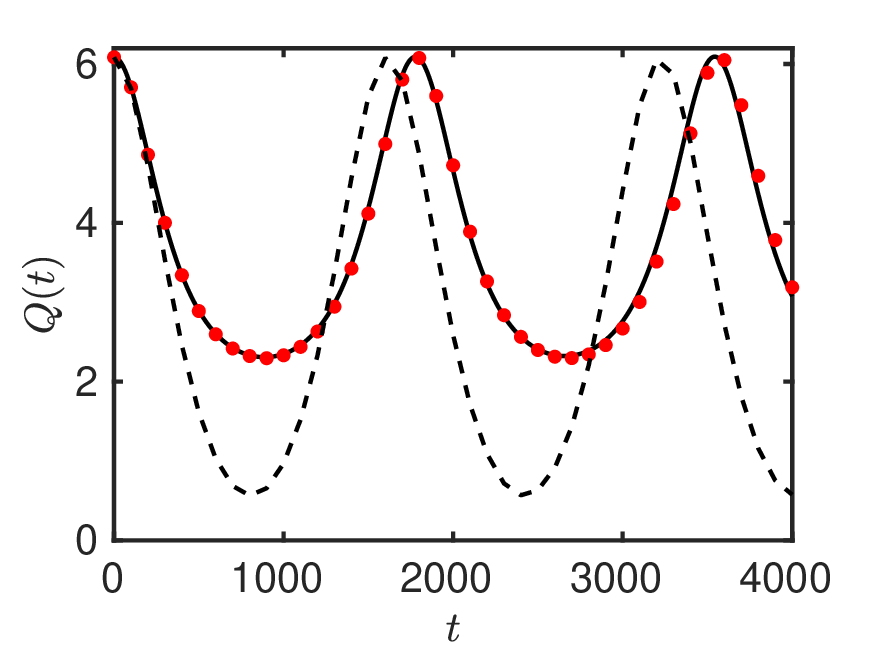} \\
		\includegraphics[width=0.4\linewidth]{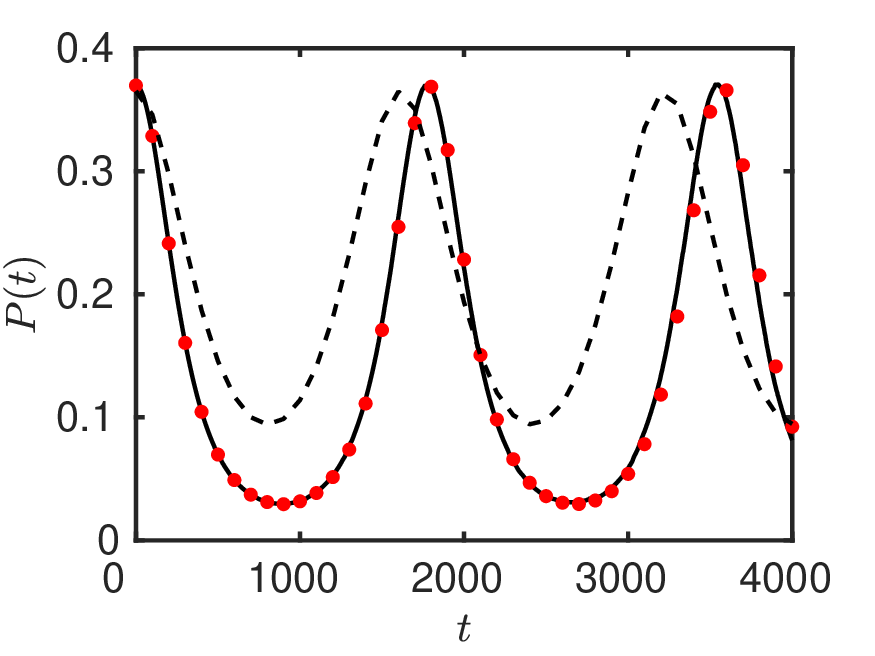} &  \includegraphics[width=0.4\linewidth]{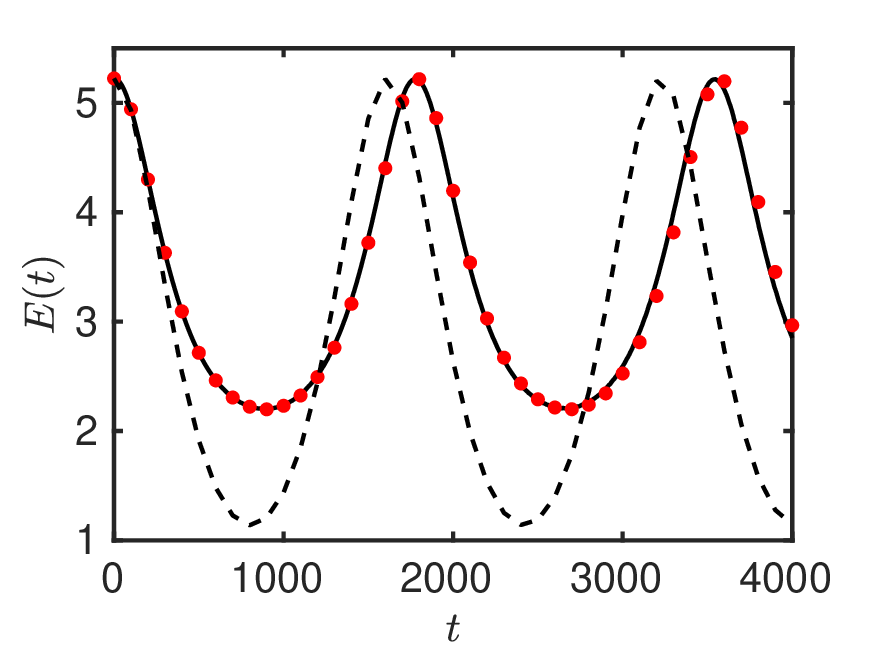}
	\end{tabular}
	\caption{
		Soliton dynamics for a lower frequency
                $\omega(0)=0.74$. The rest of parameters are the same as in
                Fig.~\ref{fig3}. 
	}
	\label{newfig}
\end{figure}

For the sake of completeness, we have also studied
  whether our variational approach for lower frequencies agrees with simulations of the full
partial differential equations for the spinor components. Specifically, for  $\omega(0)=0.74$ and the same
parameters as in Fig.~\ref{fig3}, the results obtained
are summarised in Fig. ~\ref{newfig}. Soliton dynamics is very
similar to that presented in Fig.~\ref{fig3} and the collective
coordinate theory continues  to agree very well
simulations. Certainly, the soliton moves slower and its charge is
clearly larger, but the main features of its dynamics are again
unidirectional motion, together with  oscillations of its charge,
momentum  and energy. It is also sufficiently  robust to survive for long times.

Here we recover an empirical stability criterion proposed in \cite{mertens:2011}
for the nonlinear Schr\"odinger equation that can be extended to the
nonlinear Dirac equation. It claims that if the normalized momentum $\tilde{P}=P/Q$
curve as a function of $\dot{q}$ has a negative slope, that is, 
\begin{equation}
	\label{pvmome}
	\frac{\partial \tilde{P}}{\partial \dot{q}}<0
\end{equation}
then the soliton is unstable (sufficient condition for
instability). Otherwise, the soliton can be either stable or unstable.
Note that the original $\mathcal{PT}$-symmetric ABS model (with one
  gaining mode  and one losing mode at an equal constant rate) has the
  important property that the $\mathcal{PT}$-symmetric terms preserve the Lorentz
  invariance of the Dirac equation. An exact (although not explicit)
  stationary soliton solution of that model was  found, and it turned out to be stable in most of its domain of
existence. Thus,  in that case, one can immediately obtain a family of stable travelling solitons by applying a Lorentz
boost. Remarkably, their stability does not depend on their velocity.
In  contrast, the $\mathcal{PT}$-symmetric potential considered in our paper (in
which each spinor component has gain and loss depending on $x$) breaks
the Lorentz invariance. As a consequence,  determining the stability of the travelling solitons
is a nontrivial issue, since it has to be examined for each value of the velocity.

\begin{figure}[h!]
    \centering
    \begin{tabular}{cc}
        \includegraphics[width=0.4\linewidth]{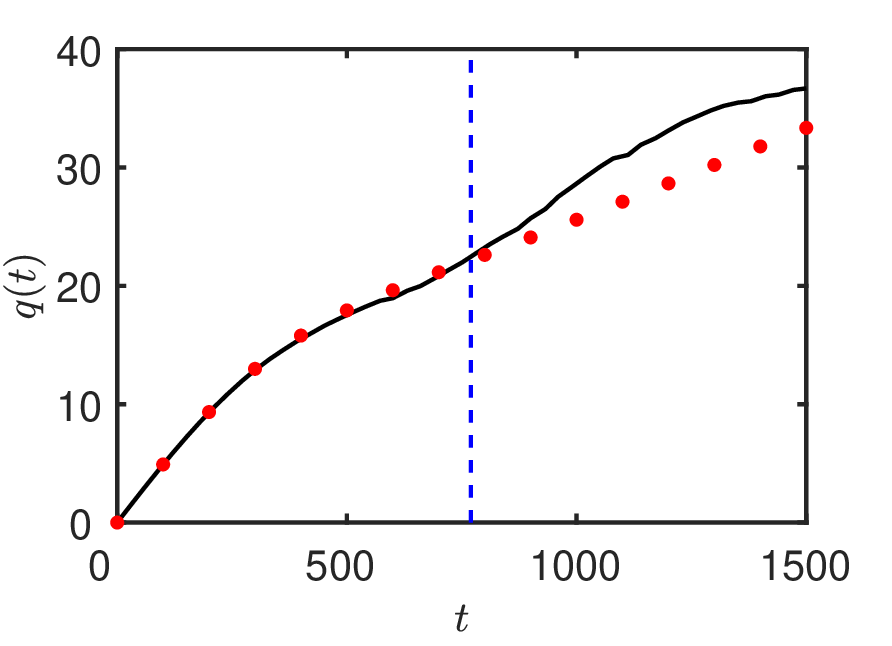} &  \includegraphics[width=0.4\linewidth]{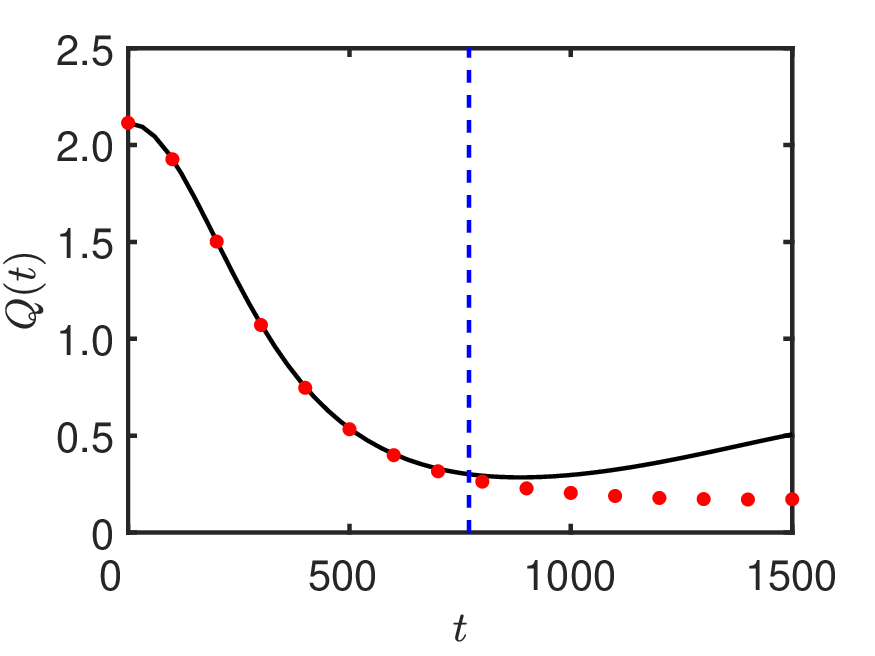} \\
        \includegraphics[width=0.4\linewidth]{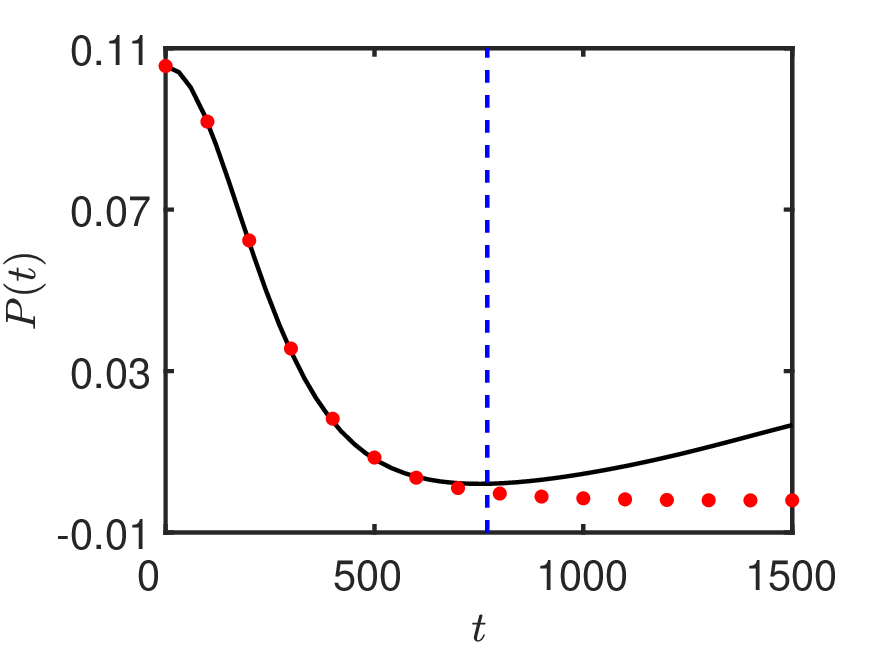} &  \includegraphics[width=0.4\linewidth]{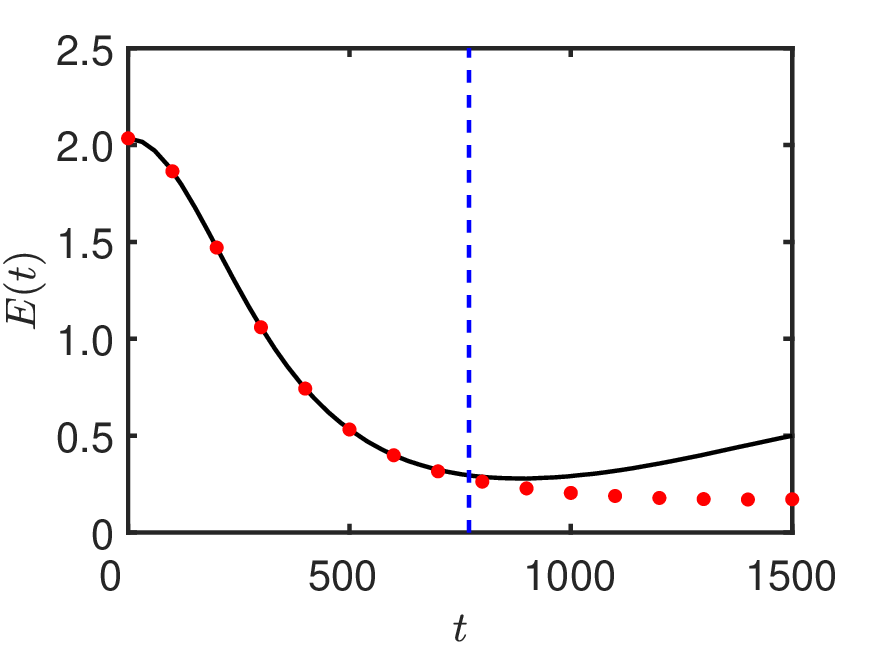} \\
        \includegraphics[width=0.4\linewidth]{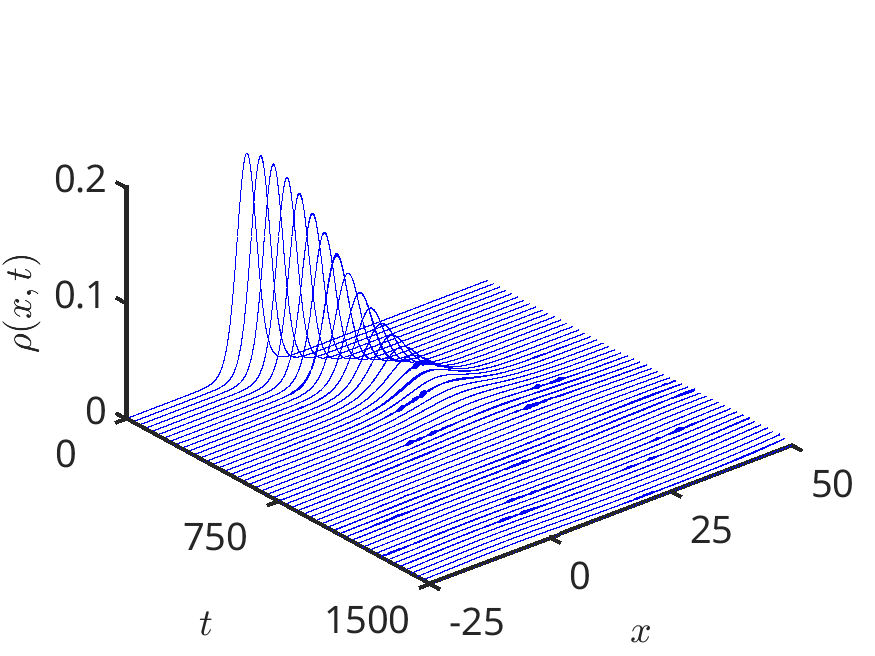} & \includegraphics[width=0.4\linewidth]{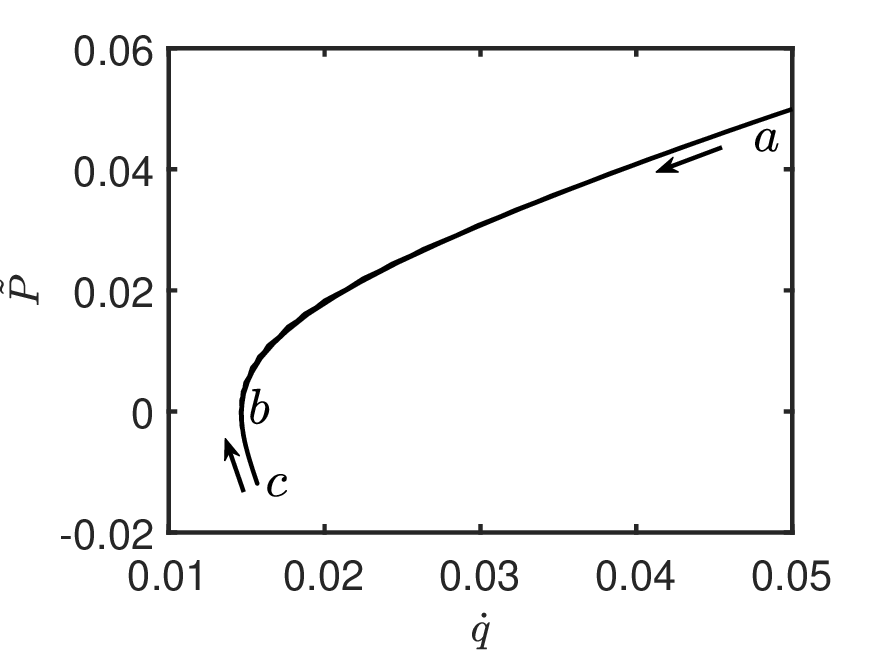}
    \end{tabular}
    \caption{Soliton dynamics for $V_0=0.001$, $W_0=0.002$, and
      $k=l=\pi/32$ (point C in Fig.\ \ref{fig1}).  In the top and middle
      panels, simulation results for the soliton position $q(t)$,
      momentum $P(t)$, charge $Q(t)$, and energy $E(t)$  (solid
      lines) are compared with  CC results (red points). The dashed
      vertical line at $t_b=770$ corresponds to point $b$ in the bottom right-hand 
      panel at which the curve $\tilde{P}(\dot{q})$ changes
      its curvature.   The bottom left-hand panel presents the
                evolution in time of the charge density.
        Initial conditions:  
        $q(0)=0$, $\omega(0)=0.9$, $\phi(0)=0$, and $\dot{q}(0)=0.05 \gtrapprox
        v_c$.  
         }
    \label{fig4}
\end{figure}

\begin{figure}[h!]
    \centering
    \begin{tabular}{cc}
        \includegraphics[width=0.4\linewidth]{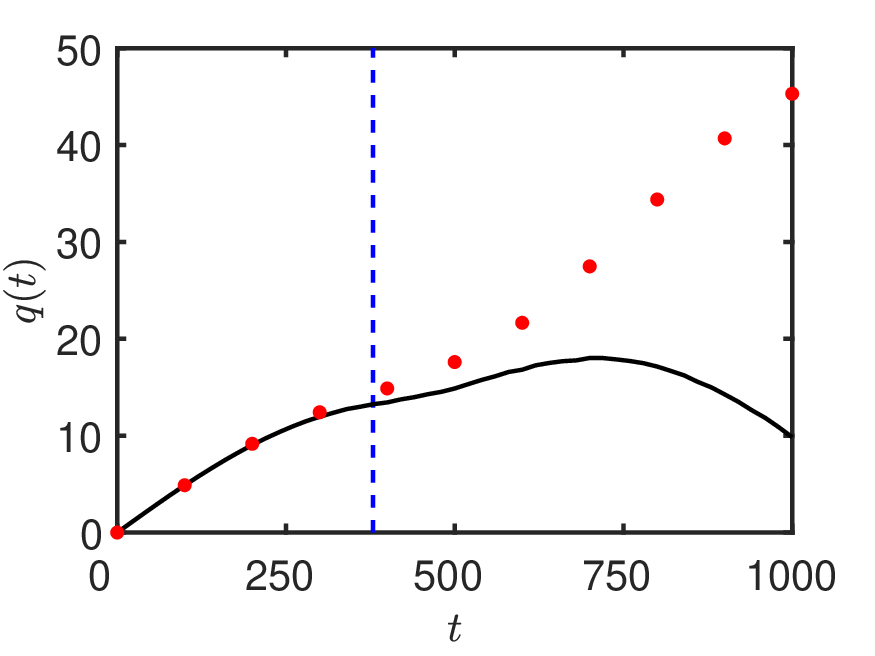} &  \includegraphics[width=0.4\linewidth]{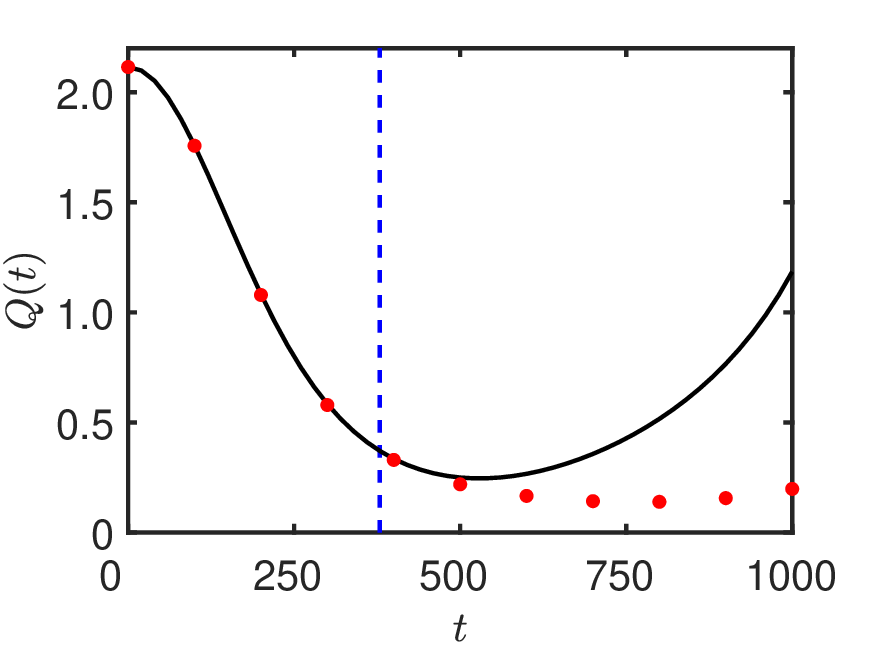} \\
        \includegraphics[width=0.4\linewidth]{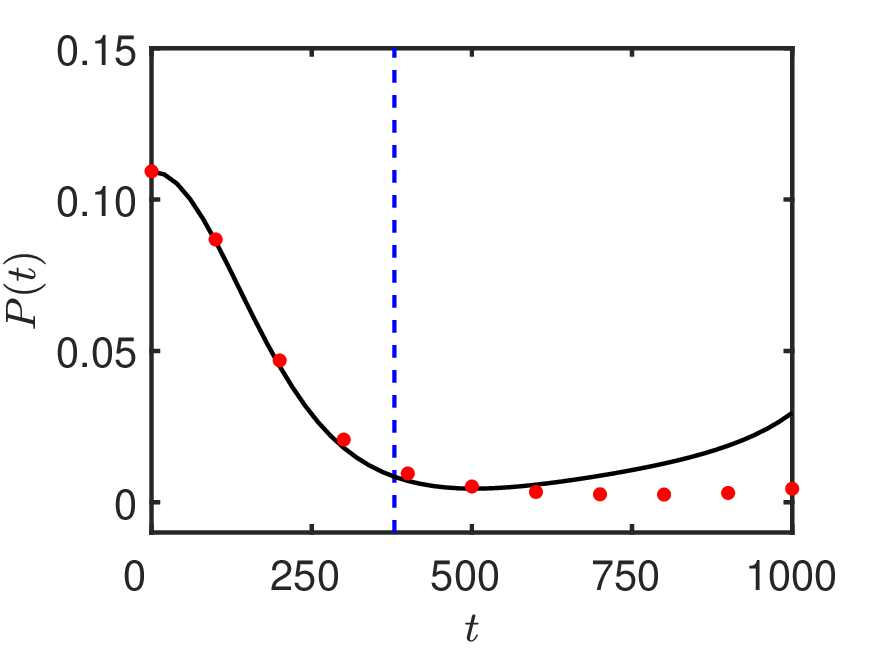} &  \includegraphics[width=0.4\linewidth]{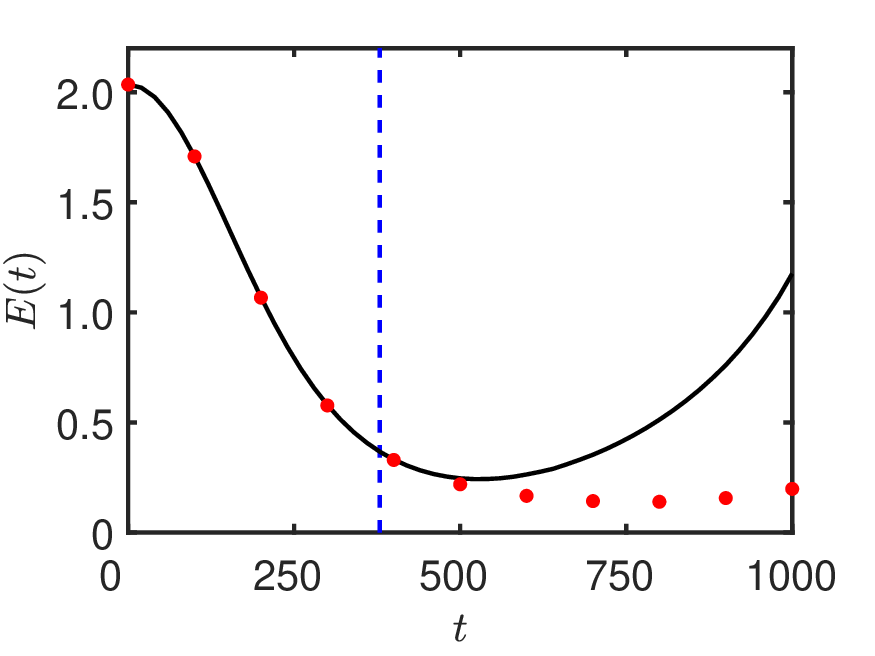} \\
        \includegraphics[width=0.4\linewidth]{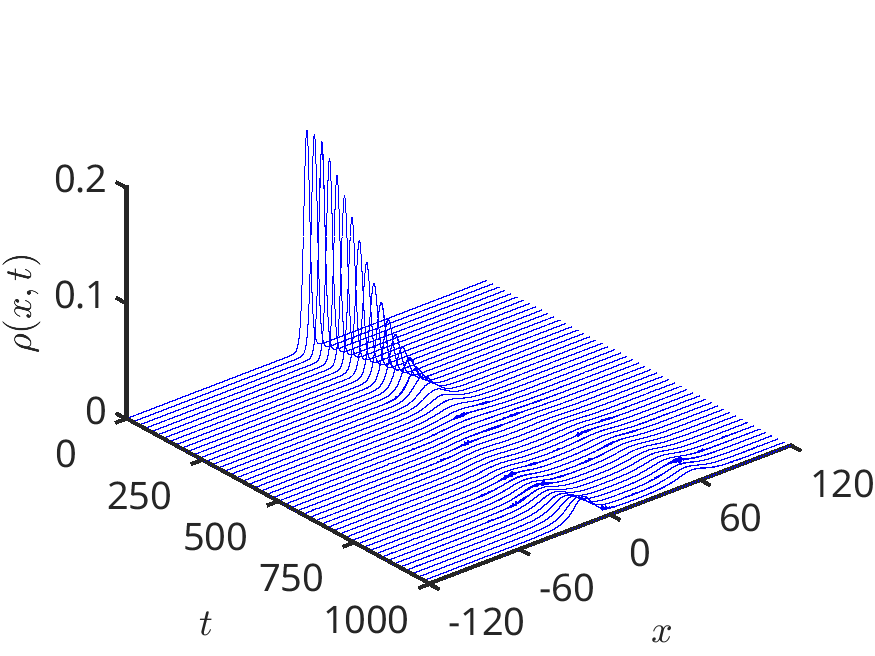} & \includegraphics[width=0.4\linewidth]{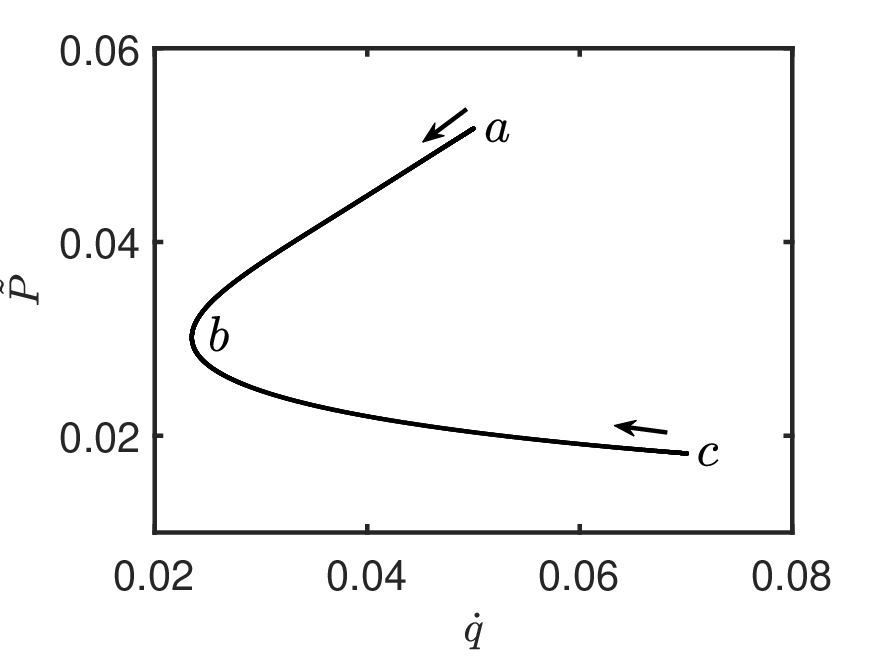}
    \end{tabular}
    \caption{Soliton dynamics for $V_0=0.001$, $W_0=0.004$, and
      $k=l=\pi/32$ (point D in Fig.\ \ref{fig1}).  In the top and middle
      panels, simulation results for the soliton position $q(t)$,
      momentum $P(t)$, charge $Q(t)$, and energy $E(t)$  (solid
      lines) are compared with  CC results (red points). The dashed
      vertical line at $t_b=379$ corresponds to point $b$ in the 
      bottom panel right-hand at which the curve $\tilde{P}(\dot{q})$ changes
      its curvature. 
      The bottom left-hand panel presents the
      evolution in time of the charge density. 
        Initial conditions:
        $q(0)=0$, $\omega(0)=0.9$, $\phi(0)=0$, and $\dot{q}(0) =0.05 \gtrapprox
        v_c$. }
    \label{fig5}
\end{figure}

For the parameters of Figs.\ \ref{fig2}-\ref{fig3} corresponding to
the dashed line between A and B in Fig.\ \ref{fig1}, 
the  $\tilde{P}(\dot{q})$  curve shows a positive slope, which is only
a necessary condition for stability. Therefore, in accordance with the
empirical stability criterion, we cannot predict the soliton
stability.  Instead, the shaded region in  Fig.\ \ref{fig1}
corresponds to unstable predictions, that is, to curves
$\tilde{P}(\dot{q})$ that have a branch with negative slope. In
Fig.\ \ref{fig4}, we show soliton dynamics, and also the
$\tilde{P}(\dot{q})$ curve (see  bottom right-hand panel) corresponding to
point C in Fig.\ \ref{fig1}, located at the border of the shaded
region.  The strong decrease of the momentum, charge and energy is
striking and can be seen as a signal of instability.  It is interesting to note that,  at $t=0$, the
$\tilde{P}(\dot{q})$ curve starts at  point $a$. As time progresses, the
values of the momentum and the velocity decrease until they reach point
$b$ at  $t_b=770$ (blue dashed vertical line in the top and middle panels). For this time interval, the CC  and simulation results agree.  
If we further increase the time, the slope of the curve becomes
negative, and, at $t_c=1414$, the minimum value of $\tilde{P}$ is reached
at point $c$ and the CC theory fails. For longer times, the points
of the curve are traversed  in the reverse direction back to  the
starting point, and so on. The soliton becomes unstable as we can clearly
observe in the bottom left-hand panel, where we have plotted the
evolution  of the 
soliton profile.
Further increments of the amplitude of the imaginary part of the
potential reinforce those effects and make them appear at earlier times.  For instance, for 
$W_0=0.004$ (point D in  in Fig.\ \ref{fig1}), one can see in the bottom right-hand panel of  Fig. \ref{fig5}
how the $\tilde{P}(\dot{q})$ curve develops a larger  branch with
negative slope. Point $b$ is now reached at $t_b=379$ and point $c$
at $t_c=766$. Instabilities appear sooner as can be verified by
observing the evolution of the charge density in the bottom left-hand panel.  

\subsection{Case $l \ne k$}

\begin{figure}[h!]
	\centering
	\begin{tabular}{cc}
			\includegraphics[width=0.4\linewidth]{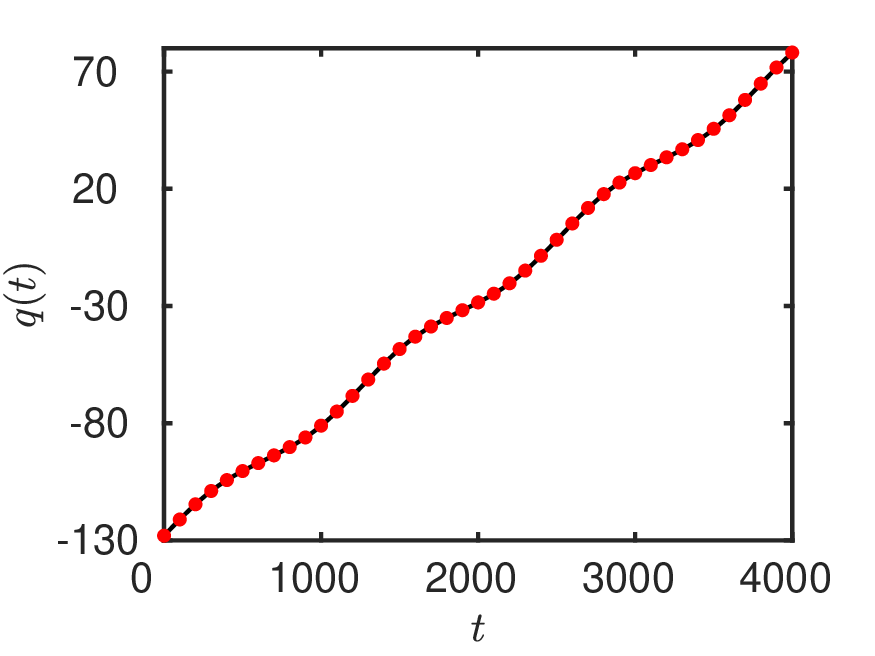} &     \includegraphics[width=0.4\linewidth]{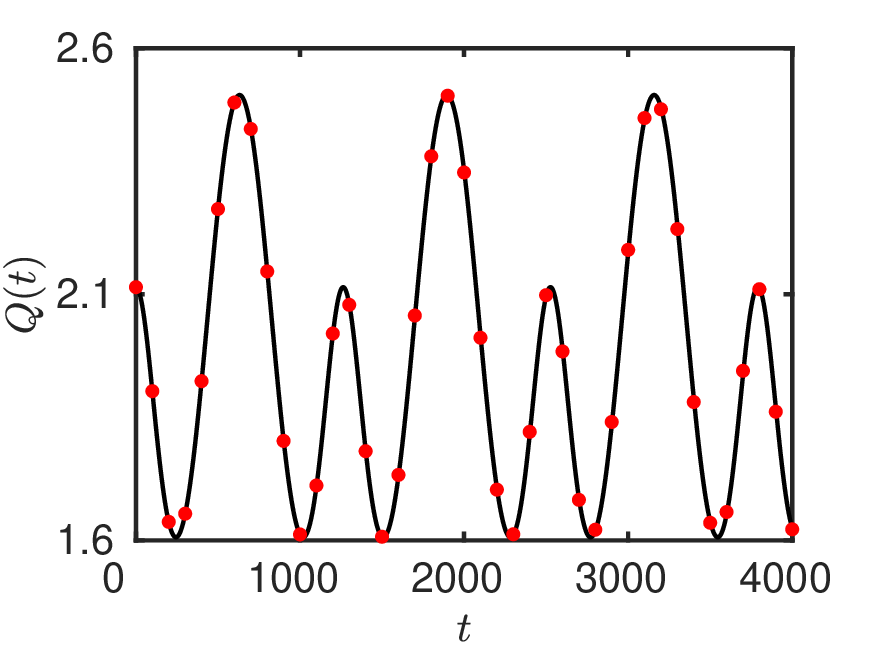} \\
			\includegraphics[width=0.4\linewidth]{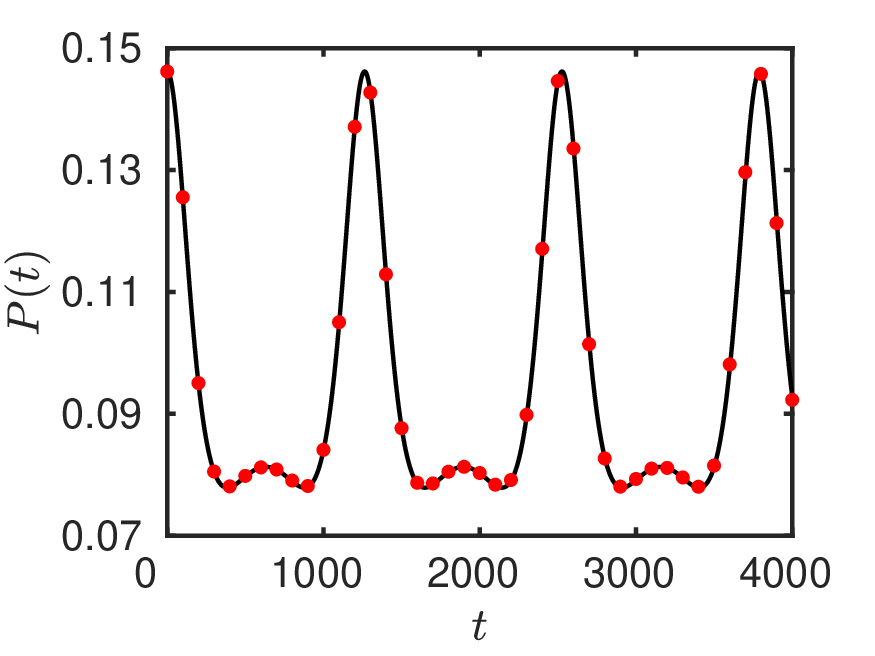} &  \includegraphics[width=0.4\linewidth]{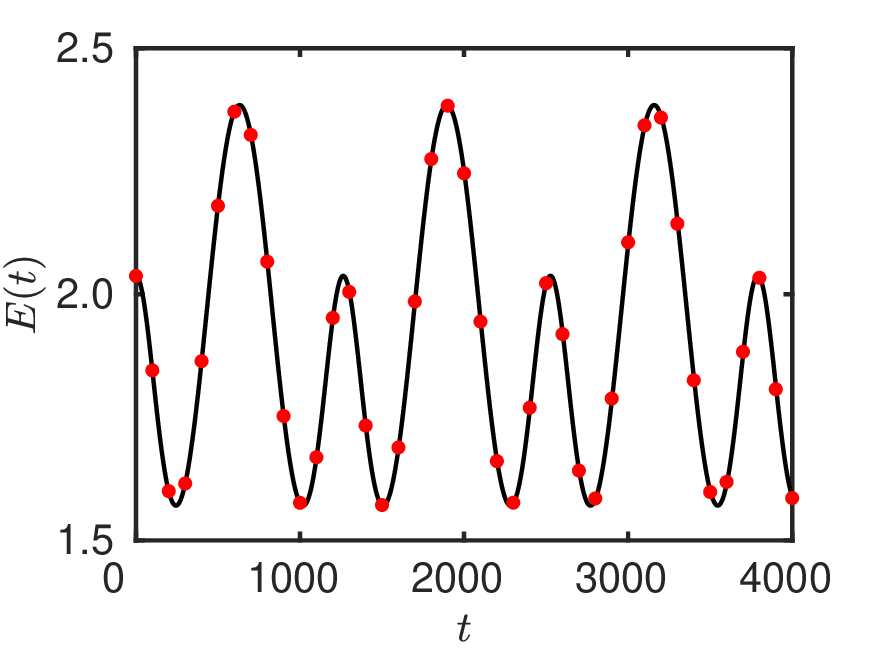} \\
		 \includegraphics[width=0.4\linewidth]{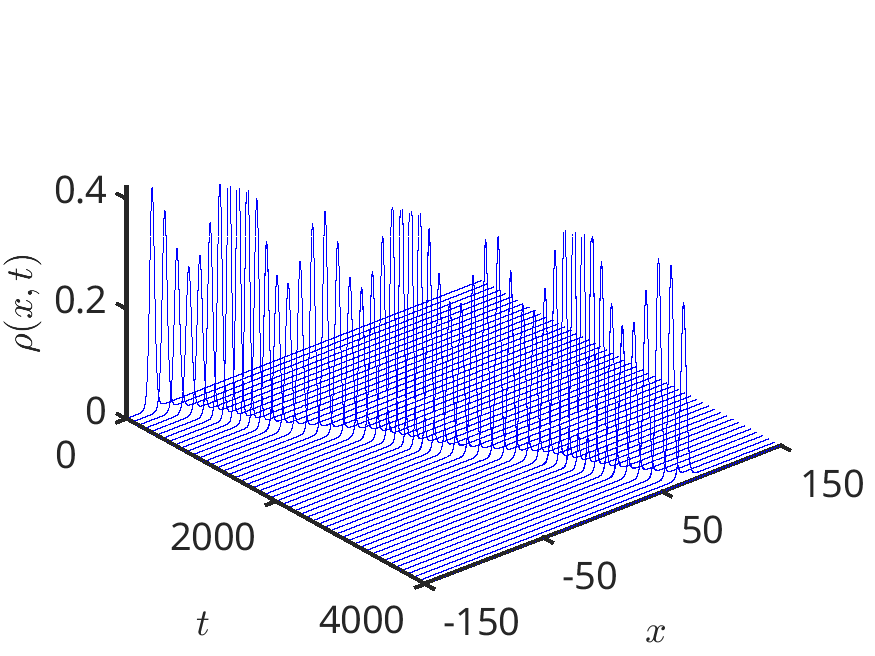}  & \includegraphics[width=0.4\linewidth]{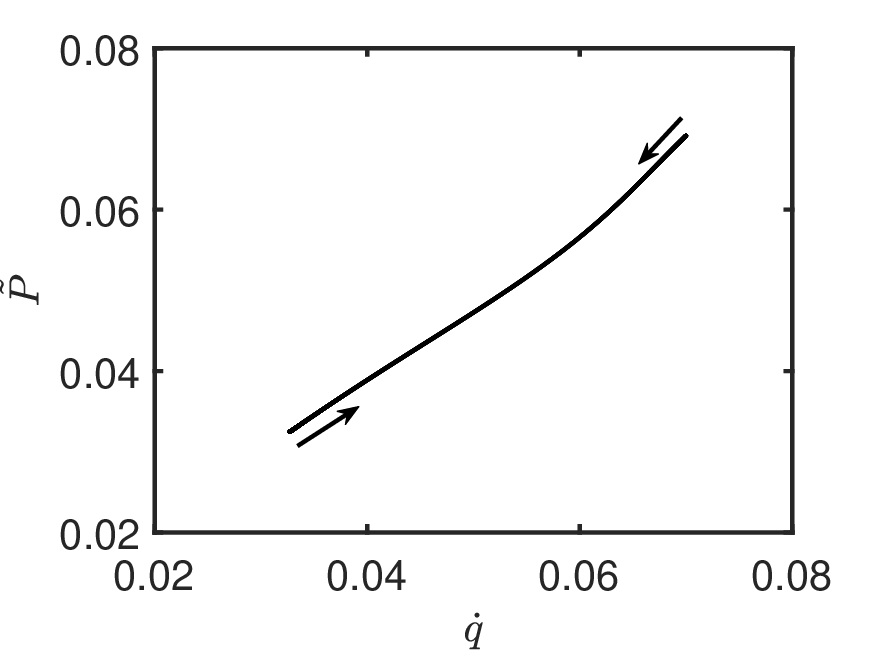} 
	\end{tabular}
	\caption{
		Soliton dynamics with the complex potential
                (\ref{potV})-(\ref{potW}) for  $V_0=W_0=0.001$,
                $k=\pi/32$, and $l=2\,k$. In the top and middle
                panels, simulation results for the soliton position
                $q(t)$, momentum $P(t)$, charge $Q(t)$, and energy
                $E(t)$  (solid lines) are compared with  CC results
                (red points). The bottom left-hand panel presents the
                evolution in time of the charge density.
		Initial conditions: $\omega(0)=0.9$, $q(0)=-128$, $\phi(0)=0$, and $\dot{q}(0)=0.07$.
	}
	\label{fig6}
\end{figure}

In  Fig. \ref{fig6}, we show soliton dynamics in a case in which the
wave numbers $k$ and $l$ of the real and imaginary part of the
potential  (\ref{potV})-(\ref{potW}) differ, specifically
$k=\pi/32$ and $l=2 k$. The CC theory (red points) continues providing a very good
approximation to simulation results.  Notice particularly the appearance of two
frequencies in the oscillations of the charge and of the energy, which
carries on oscillating in phase. The stability curve
$\tilde{P}(\dot{q})$ (bottom right-hand panel) displays a positive slope at
all times and, in fact, in the  bottom left-hand panel, one can see  that the
soliton survives without distortions for significantly long times.

\begin{figure}[h!]
	\centering
	\begin{tabular}{cc}
		\includegraphics[width=0.4\linewidth]{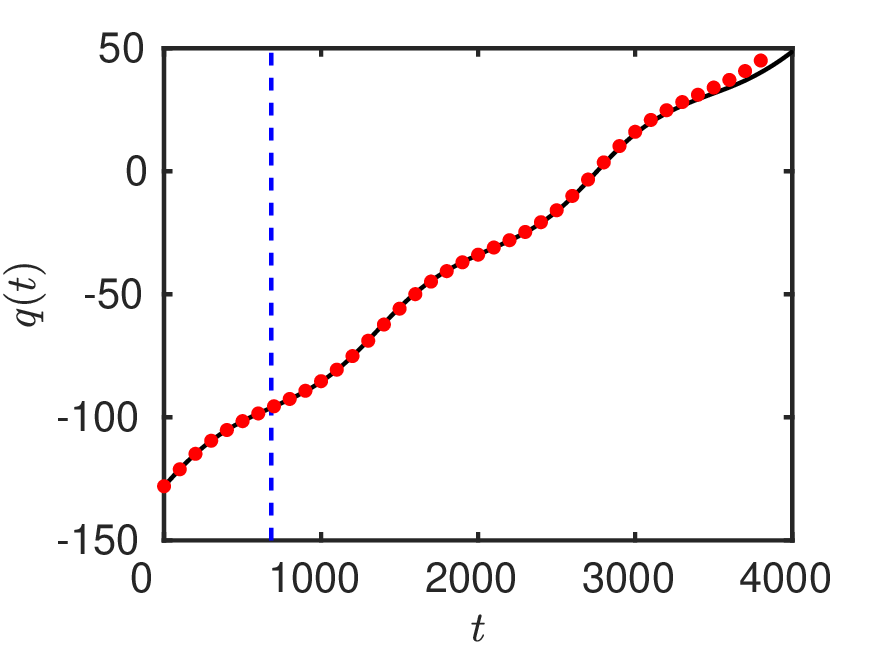} &     \includegraphics[width=0.4\linewidth]{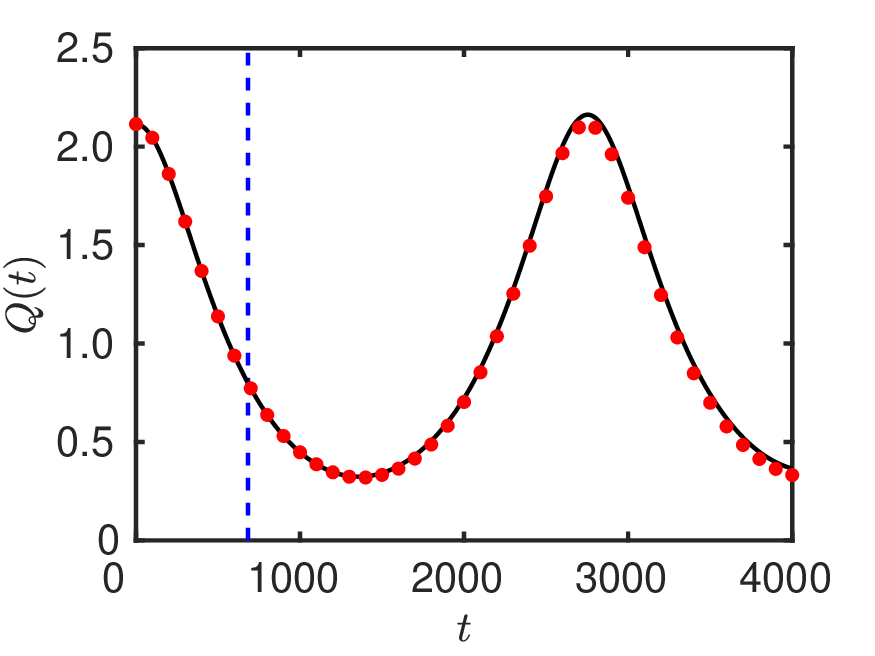} \\
		\includegraphics[width=0.4\linewidth]{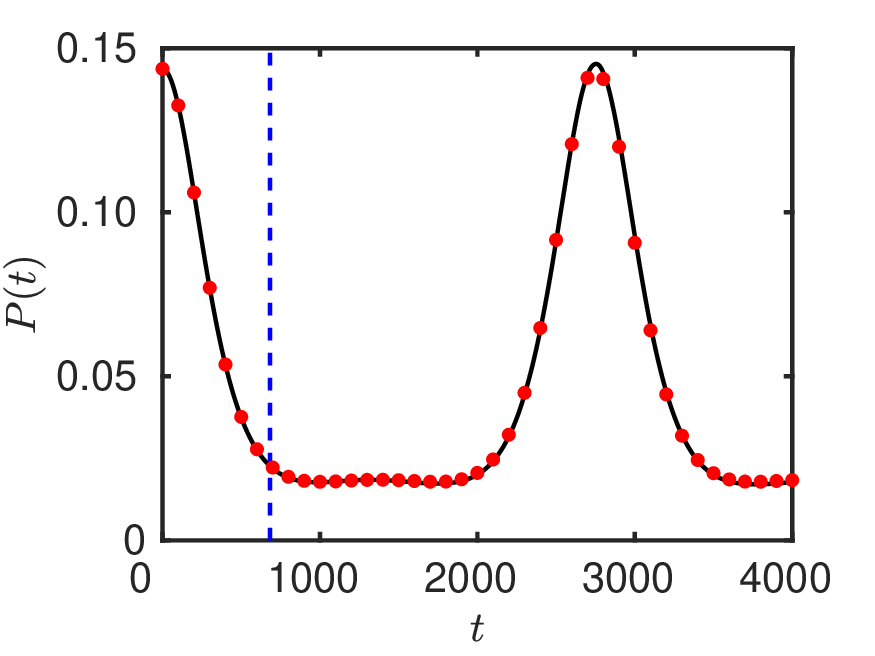} &  \includegraphics[width=0.4\linewidth]{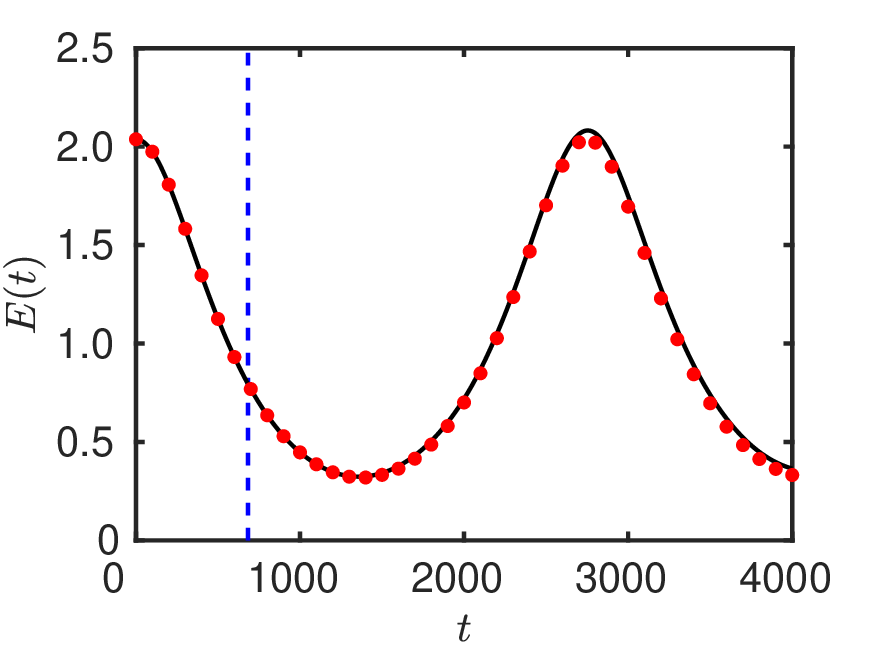} \\
		\includegraphics[width=0.4\linewidth]{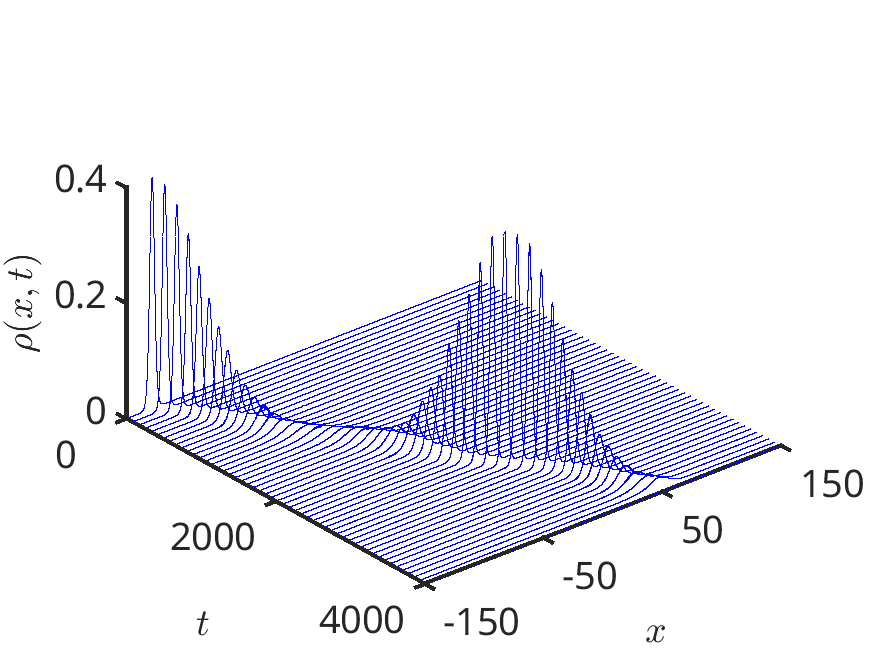}  & \includegraphics[width=0.4\linewidth]{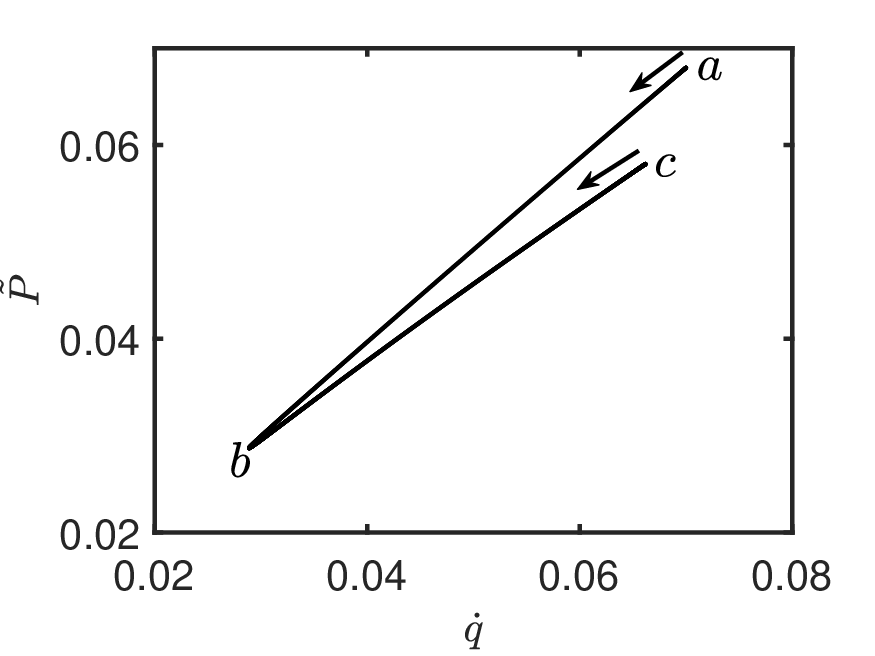} 
\end{tabular}
\caption{
Soliton dynamics with the complex potential (\ref{potV})-(\ref{potW}) for  $V_0=W_0=0.001$,  $k=\pi/32$, and $l=k/2$. In the top and middle 	panels, simulation results for the soliton position $q(t)$, momentum $P(t)$, charge $Q(t)$, and energy $E(t)$  (solid lines) are compared with  CC results (red points). The dashed vertical line at $t_b=683$ corresponds to point $b$ in the 
bottom right-hand panel at which the curve $\tilde{P}(\dot{q})$ has a
cusp. The bottom left-hand panel presents the evolution in time of the
soliton profile.
Initial conditions: $\omega(0)=0.9$, $q(0)=-128$, $\phi(0)=0$, and $\dot{q}(0)=0.07$.
}
	\label{fig7}
\end{figure}

In contrast with the case $l=2k$, soliton dynamics for $l=k/2$ 
exhibits only one frequency as can be
observed in  Fig. \ref{fig7}. Interestingly, the curve
$\tilde{P}(\dot{q})$ shows a cusp at point $b\, (t_b=683)$, which
corresponds to the dashed vertical lines in the top and middle
panels. Note the surprisingly large amplitudes of charge and energy
oscillations. Despite the fact that they are reduced up to approximately 25$\%$ of
their initial values,  the soliton is a long-life structure, as can be
observed in the 
bottom left-hand panel.  

\section{Conclusions} \label{sec5}
Motivated by a burst of intense research activity  on
$\cal{PT}$-symmetric systems, triggered by the seminal contribution of
Bender and Boettcher \cite{bender:1998}, we have
investigated soliton dynamics in the $\cal{PT}$-symmetric ABS
model with a complex potential. This model belongs to the general family of the nonlinear Dirac equation. The evolution equation for the
Dirac spinor is derived from a Lagrangian density, where the real part
$V(x)$  of
the potential is introduced in the standard fashion through the
gauge covariant derivative, and the imaginary part $W(x)$ of the potential
is incorporated by means of a dissipation function. In order to preserve the $\cal{PT}$-symmetry, $V(x)$ and $W(x)$ must be even and odd functions, respectively.

For the specific choice of periodic functions (\ref{potV})-(\ref{potW}), we have developed a collective coordinate
theory for the soliton dynamics using an ansatz with five collective
coordinates: the position, the rapidity, the frequency, the
momentum, and the phase. This approach allows us to obtain the
Lagrangian as a function of these variables and theirs time
derivatives. The evolution of these collective coordinates is governed
by four coupled nonlinear ordinary differential equations and one
algebraic equation.  

We have focused on  the influence of the imaginary part of the potential
on soliton dynamics.  Approximate analytical expressions for
the charge and the energy in the non-relativistic regime have been derived by considering 
small values of the amplitude  
$W_0$, and expanding  the collective
variables in powers of $W_0$. We conclude that the main effect of the imaginary part of
the potential is to induce oscillations in the charge and in the
energy, which are constant at zero order.  At first order, it is found that both oscillate in phase and with the same frequency as the momentum.  Such frequency is captured by the zero-order approximation for the momentum, and depends on whether the soliton initial velocity
$\dot{q}(0)$ is below or above a critical velocity $v_c$, which acts as a separatrix between   
 oscillatory  and unbounded motion, respectively.
Comparison of the predictions of the  CC theory with numerical simulations of the
exact spinor evolution equation shows an excellent
agreement. As expected, the analytical approximations  work reasonably only for small values of $W_0$ and $\dot{q}(0)$.  
  
We have numerically explored the dependence of $v_c$ on $W_0$, and observed the emergence of instabilities as $W_0$ increases. 
At this point, we have recovered and extended to the nonlinear Dirac equation an empirical stability criterion that had already been employed for the nonlinear Schrödinger equation. The criterion claims that  soliton instability appears when the normalized momentum $\tilde{P}(\dot{q})$ has a negative slope. We have identified a region in the parameter plane $(\dot{q}(0), W_0)$ where $\tilde{P}(\dot{q})$ develops a branch with a negative slope, and have verified the criterion by studying soliton evolution through numerical simulations of the spinor equations. Interestingly, the CC theory works very well up to the point at which the slope of $\tilde{P}(\dot{q})$ becomes negative. From this point onwards, the CC theory fails. This effect is reinforced and the soliton lifetime is shorter while $W_0$ is increased.

Finally, we have also investigated the influence on the dynamics of
setting different wave numbers in the real and imaginary part of the
complex potential. Apart from detecting the appearance of multiple
frequencies in the charge and  energy oscillations, the most striking
observation is that of the long soliton lifetime even when charge and
energy oscillations have a very large amplitude.

An interesting perspective for future work, pointed out by
  the referees, is  to explore
  soliton dynamics for other choices of the complex potential. For
  instance, one can think about  a
complex potential which presents twofold $\mathcal{PT}$-symmetry as in
Ref. \cite{cole:2016}. Another possibility is to study what happens
when the periodic distribution of the imaginary potential in the
two spinor components are in opposite to that of each other. These
lines of research would certainly be worth investigating.

\section*{Acknowledgments}
B.S-R. acknowledges financial support from the Ministerio de Ciencia e
Innovación through PID2021-122588NBI00 funded by MCIN/AEI/10.13039/501100011033/ 
and from Grant ProyExcel
00796 funded by Junta de Andaluc\'{\i}a’s PAIDI 2020
programme.
N.R.Q.  acknowledges financial support from the Spanish projects PID2020-113390GB-I00 (MICIN), PY20$\_$00082
(Junta de Andalucia).


\section*{Data availability statement}
The data that support the findings of this study are available upon request from the authors.

\section*{ORCID iDs}
Franz G Mertens https://orcid.org/0000-0002-0574-2279 \\
Bernardo Sánchez-Rey https://orcid.org/0000-0003-3170-154X\\
Niurka R Quintero https://orcid.org/0000-0003-3503-3040

\section*{References}

\bibliographystyle{unsrt}
\bibliography{dirac3chaos}

\end{document}